\documentclass[aps,prb,twocolumn,longbibliography]{revtex4-2}
\usepackage[T1]{fontenc}
\usepackage[utf8]{inputenc}
\usepackage{amsmath,amssymb,bm,graphicx,hyperref}
\usepackage{array,newunicodechar}
\newunicodechar{λ}{\ensuremath{\lambda}}
\newunicodechar{κ}{\ensuremath{\kappa}}
\newunicodechar{η}{\ensuremath{\eta}}
\newunicodechar{γ}{\ensuremath{\gamma}}
\newunicodechar{τ}{\ensuremath{\tau}}
\newunicodechar{ε}{\ensuremath{\varepsilon}}
\newunicodechar{ν}{\ensuremath{\nu}}
\newunicodechar{ξ}{\ensuremath{\xi}}
\newunicodechar{∂}{\ensuremath{\partial}}
\newunicodechar{⊥}{\ensuremath{\perp}}
\newunicodechar{≃}{\ensuremath{\simeq}}
\newunicodechar{≡}{\ensuremath{\equiv}}
\newunicodechar{∝}{\ensuremath{\propto}}
\newunicodechar{∑}{\ensuremath{\sum}}
\newunicodechar{δ}{\ensuremath{\delta}}
\newunicodechar{±}{\ensuremath{\pm}}
\newunicodechar{⟨}{\ensuremath{\langle}}
\newunicodechar{⟩}{\ensuremath{\rangle}}
\newunicodechar{ℰ}{\ensuremath{\mathcal E}}

\begin{document}
\title{Rotational Perturbation Theory for Magnetic Interactions}
\author{Vladimir Antropov}
\affiliation{Ames National Laboratory, Ames, Iowa 50011, USA}
\begin{abstract}
We develop rotational perturbation theory (RPT) by distinguishing two magnetic energy curvatures. Prescribed rotations of the magnetic potential give the field curvature $H^B$, whereas constrained rotations of magnetic moments, with the electronic state allowed to respond, give the moment curvature $H^M$ through the inverse static susceptibility. These curvatures generally differ and need not yield the same exchange interactions. The prescribed-field expansion recovers conventional exchange expressions without assuming a Heisenberg model. For noncollinear states, we extract the isotropic Heisenberg contribution and identify the part it does not describe. We derive planar and conical spin-spiral expressions, including contributions from nonlocal magnetic potentials. We derive the long-wavelength ferromagnetic stiffness and two-sublattice antiferromagnetic spin-wave velocity, specifying the dynamical assumptions needed to relate them to static curvatures. For correlated electrons, we identify prescribed self-energy rotations and distinguish their trace-logarithm contribution from the additional terms required for a conserving response.
\end{abstract}
\maketitle
\section{INTRODUCTION}

Magnetic interactions can be calculated from changes in electronic energy caused by rotations of local magnetic directions. The result depends on whether exchange fields or moments are rotated, how the electrons respond, and how the energy is mapped onto a spin model. These distinctions are particularly important in itinerant magnets. Our central question is how the energy curvature changes when we rotate the magnetic potential rather than constrain the magnetic moments.

In the field approach, the exchange-field directions are fixed in the one-electron Hamiltonian. The resulting energy derivatives give the field curvature $H^B$, from which exchange coefficients $J_{ij}^{B}$ can be obtained by comparison with a bilinear spin model. For moment rotations, the directions of the magnetic moments are constrained while the remaining electronic degrees of freedom relax. The curvature $H^M$ of the constrained total energy is determined by the inverse static magnetic susceptibility and can similarly be used to define $J_{ij}^{M}$. The two curvatures generally differ. The magnetic local force theorem (LFT) and the Harris--Foulkes identity concern different energy variations. The LFT concerns first-order energy changes about a stationary state. For rigid input rotations in the local spin-density approximation (LSDA), the Harris--Foulkes identity relates the field curvature to a total-energy expression. Neither result makes $H^B$ equal to $H^M$. Their leading long-wavelength terms agree only under restricted assumptions.

Another question is how electronic energy changes determine spin-model parameters. For a bilinear Heisenberg model, the transverse energy curvature is determined by pair exchange coefficients $J_{ij}$. Biquadratic, multispin, and other non-Heisenberg interactions also contribute to the energy change. Exchange coefficients inferred from small rotations can therefore depend on the reference state and need not describe other configurations.

Nonlocal magnetic potentials introduce another complication. Many exchange expressions use local exchange-correlation fields, although the resulting couplings may extend over many coordination shells. Orbital-dependent functionals, nonlocal self-energies, and gradient-dependent terms require the corresponding rotational vertices and electronic response.

The distinction also matters for weak or poorly defined local moments. In magnets with well-developed moments, transverse rotations provide natural low-energy coordinates. In itinerant systems, longitudinal fluctuations can couple to transverse rotations. Magnetic correlations can also persist without a static reference moment. In that case no equilibrium moment direction exists, and spin-density fluctuations must be described through the susceptibility.

Finally, static interactions must be connected with spin dynamics. Spin-wave frequencies require the dynamical response as well as the static curvature. In a ferromagnet with $\omega\propto q^2$, the frequency-linear and $q^2$ terms are equally important; antiferromagnetic spin-wave velocities likewise require a dynamical model.

Microscopic exchange theories initially emphasized the static electronic response to magnetic perturbations. Ruderman and Kittel \cite{ref1}, Kasuya \cite{ref2}, and Yosida \cite{ref3} established the theory of indirect exchange mediated by conduction electrons. Freeman and collaborators subsequently employed realistic band theory to investigate conduction-electron polarization and magnetic coupling through the static susceptibility of a nonmagnetic reference state \cite{ref4,ref5}. Liu \cite{ref6} extended the Ruderman--Kittel--Kasuya--Yosida approach to spin-polarized electronic systems using a Hubbard-like model, while Cooke \cite{ref7} developed related band-structure formulations.

Korenman, Murray, and Prange developed the local-band theory of itinerant ferromagnetism \cite{ref8,ref9,ref10,ref11,ref12}, in which the electronic structure remains locally spin polarized while the magnetization direction varies slowly in space and time. They treated rotations perturbatively about a magnetized state and investigated spin-wave dynamics, nonlinear magnetic fluctuations, and thermodynamics. In these calculations, the electronic perturbations rotated local exchange fields, even when described in terms of magnetic moments. Prange and Korenman (PK) subsequently employed a Hubbard--Stratonovich formulation to obtain an effective classical Heisenberg model \cite{ref10,ref11}. Wang and collaborators \cite{ref13} later applied this approach to Fe and Ni using realistic electronic bands to calculate interatomic exchange interactions.

Prescribed moment configurations also motivated constraining-field and torque methods. Edwards \cite{ref14} introduced constraining fields into calculations of exchange interactions $J_{ij}$, allowing magnetic moments to be held along prescribed directions that need not be stationary configurations of the unconstrained electronic system. In the same year, Small and Heine \cite{ref15} developed a complementary torque-based approach within electronic theory, deriving the torque acting on an atomic moment from both the electronic equation of motion and the total energy. They showed how the resulting torques could determine interatomic magnetic couplings for arbitrary moment configurations and applied the method to bcc Fe.

A further development was the multiple-scattering formulation of magnetic exchange introduced by Liechtenstein, Katsnelson, and Gubanov (LKG) \cite{ref16,ref17}. They obtained interatomic exchange coefficients from infinitesimal rotations of single-site scattering matrices ($t$ matrices), interpreting the resulting interactions as those between rotated magnetic moments. The single-site $t$ matrix represents scattering from the potential inside an atomic sphere; its spin-dependent part is not simply an integrated exchange-correlation field. Rotating this matrix corresponds to a prescribed rotation of the magnetic scattering potential, not to a constraint on the integrated atomic moment. Their exchange expressions were similar to those of PK and were suited to multiple-scattering calculations.

LKG and collaborators also formulated the magnetic LFT \cite{ref18}, relating first-order total-energy variations to frozen-potential electronic variations at a stationary reference. Their exchange expressions were obtained from second-order frozen-potential variations. First-order stationarity alone does not establish the corresponding total-energy curvature. Subsequent extensions using the coherent potential approximation enabled the treatment of chemical and magnetic disorder, while relativistic generalizations provided access to the Dzyaloshinskii--Moriya interaction (DMI) and anisotropic exchange \cite{ref19,ref20}. Related formulations incorporating electronic self-energies and many-body effects were later developed for correlated materials \cite{ref21}.

Solovyev and Terakura \cite{refSolovyevTerakura1998} examined the relation between frozen-potential electronic energy changes and total-energy variations for small rotations in nonrelativistic density-functional theory. Their analysis used the rotational properties of the exchange-correlation functional and considered corrections from induced changes in charge density and magnetization magnitude.

An alternative approach starts from the transverse magnetic susceptibility, which describes the response of magnetic moments rather than prescribed rotations of the exchange potential. In the theory of itinerant magnetism, the exchange interactions of an effective Heisenberg model were traditionally related to the inverse static transverse susceptibility \cite{Moriya1985}, but this relation was not explicitly used in exchange calculations. Ref.~\cite{ref22} introduced inverse-susceptibility exchange into electronic-structure calculations, compared it with conventional exchange expressions, and examined the long-wavelength approximation. That work distinguished field from moment rotations, obtained exchange interactions without explicitly imposing Edwards' constraining fields, and examined the frequency-linear response of spin dynamics.

We use rotational perturbation theory (RPT) to calculate the prescribed-field curvature and compare it with the constrained-moment curvature obtained from inverse susceptibility. For local magnetic potentials, RPT reproduces the PK and LKG exchange expressions \cite{ref11,ref16}; Appendix~\ref{app:pk_lkg} compares their band and multiple-scattering forms. The analysis extends to noncollinear states, planar and conical spin spirals, nonlocal magnetic potentials, and long-wavelength spin dynamics. For correlated electrons, we distinguish prescribed rotations of the equilibrium self-energy from its induced response in a stationary calculation.

\section{GENERAL ELECTRONIC AND ROTATIONAL PERTURBATION THEORY}

For small rotations $\delta\boldsymbol\theta_i$, the rotational Hessian $H$ contains the second derivatives of the energy with respect to the rotation angles and determines the quadratic energy change,
\[
\delta E^{(2)}=\frac12\sum_{ij,ab}\delta\theta_i^a H_{ij}^{ab}\delta\theta_j^b.
\]
We use \emph{Hessian} for the matrix of second derivatives and \emph{curvature} for the corresponding second-order energy coefficient of a specified rotation.

\subsection{Magnetic representations and general electronic perturbation theory}
\label{sec:pt_repr}

The exchange field \(\mathbf{B}(\mathbf r)\) enters the Kohn--Sham or another effective electronic Hamiltonian, whereas constraints on \(\mathbf{M}(\mathbf r)\) fix the local moment directions. We write the electronic Hamiltonian as

\begin{equation}
{\widehat{H}}_{\mathrm{KS}} = {\widehat{H}}_{0}\  - \ \bm{\sigma} \cdot \mathbf{B}(\mathbf{r}) = {\widehat{H}}_{0}\  - \ \frac{1}{2}\bm{\sigma} \cdot \mathbf{\Delta}(\mathbf{r}).
\label{eq:ks_hamiltonian}
\end{equation}

In atom-centered descriptions, the integrated moment and exchange-correlation field of sphere $\Omega_i$ are
\begin{equation}
 \mathbf M_i=\int_{\Omega_i}d^3r\,\mathbf m(\mathbf r),\qquad
 \boldsymbol{\mathcal B}_i=\int_{\Omega_i}d^3r\,\mathbf B_{\rm xc}(\mathbf r).
\label{eq:sphere_integrals}
\end{equation}
For a rotationally invariant local LSDA functional, $\mathbf B_{\rm xc}(\mathbf r)$ is parallel to $\mathbf m(\mathbf r)$ pointwise. Nevertheless, when the magnetization direction and the local exchange splitting vary within a sphere, their integrals need not be parallel:
\begin{equation}
 \int_{\Omega_i}d^3r\,\mathbf m(\mathbf r)\times\mathbf B_{\rm xc}(\mathbf r)=0,
 \qquad \mathbf M_i\times\boldsymbol{\mathcal B}_i\ne0\quad\text{in general}.
\label{eq:sphere_torque}
\end{equation}
The cross product of these integrated quantities is not the exchange-correlation torque on the atomic moment. In the nonrelativistic local-LSDA spin-continuity equation, the integrated moment dynamics contain the spin-current flux through the boundary of the atomic region and any external or constraining-field torques; the local exchange-correlation torque vanishes pointwise. In stationary constrained configurations, these contributions balance. Nonlocal spin-dependent operators and spin--orbit coupling (SOC) require their corresponding additional spin-balance terms. Rotating the magnetic scattering potential does not fix the direction of the integrated atomic moment. Constrained-moment rotations instead fix that direction while allowing the electronic state to respond.

We expand the rotated Hamiltonian as \(H = H_{0}\  + \ V^{(1)}\  + \ V^{(2)}\  + \  \cdot \cdot \cdot\), with superscripts denoting orders in the rotation angle. The linear perturbation determines the first-order energy change. At second order, the energy contains both the direct second-order change of the Hamiltonian and the electronic response to the linear perturbation. We use this static expansion for spin spirals and long-wavelength modes and treat time dependence separately in Sec.~\ref{sec:pt_time}.

\subsection{Rotations of the effective magnetic field}
\label{sec:pt_field}

For a noncollinear magnetic state, the magnetic part of the one-electron Hamiltonian is

\begin{equation}
H_{mag} = - \sum_{i}^{}{b_{i}\,\mathbf{e}_{i} \cdot \bm{\sigma}}
\label{eq:orig001}
\end{equation}
where \(\mathbf e_i\) specifies the direction of the prescribed site magnetic potential in this rigid-rotation representation, and \(2b_i=\Delta_i\) is its spin splitting in the local orbital space. In a general noncollinear state, \(\mathbf e_i\) need not coincide with the direction of the integrated atomic moment. In the simplest case, \(b_i\) is a scalar. The rotations are transverse and keep the magnitude fixed,

\begin{equation}
\mathbf{e}_{i} \rightarrow \mathbf{e}_{i} + \delta\mathbf{e}_{i}
\label{eq:orig002}
\end{equation}
At a reference state stationary with respect to the unconstrained electronic total-energy functional, the first angular derivative of that total energy vanishes:

\begin{equation}
\left. \ \frac{\partial E}{\partial\theta_{i}^{a}} \right|_{0} = 0
\label{eq:orig003}
\end{equation}
A constrained state can still have a torque. We must then include the constraining fields and the geometric terms when calculating the second derivative.

At each site, the transverse plane is spanned by two orthonormal vectors \(\mathbf{u}_{i}^{1}\), \(\mathbf{u}_{i}^{2}\) perpendicular to \(\mathbf e_i\). An infinitesimal rotation is described by

\begin{equation}
\delta\mathbf{\theta}_{i} = \sum_{a = 1,2}^{}{\theta_{i}^{a}\mathbf{u}_{i}^{a}}
\label{eq:orig004}
\end{equation}
The tangent vectors are

\begin{equation}
\mathbf{t}_{i}^{a} = \mathbf{u}_{i}^{a} \times \mathbf{e}_{i}
\label{eq:orig005}
\end{equation}
so that, to first order,

\begin{equation}
\delta\mathbf{e}_{i} = \sum_{a}^{}{\theta_{i}^{a}\mathbf{t}_{i}^{a}}
\label{eq:orig006}
\end{equation}
A small tilt changes the electronic Hamiltonian. At site $i$, it is

\begin{equation}
V_{i}^{(1)} = - b_{i}\,\delta\mathbf{e}_{i} \cdot \bm{\sigma}
\label{eq:orig007}
\end{equation}
or

\begin{equation}
V_{i}^{(1)} = \sum_{a}^{}{\theta_{i}^{a}V_{i}^{a}},\quad\quad V_{i}^{a} = - b_{i}\,\mathbf{t}_{i}^{a} \cdot \bm{\sigma}
\label{eq:orig008}
\end{equation}
The rotational vertex is transverse to the local exchange field and contains the full exchange splitting. We expand in the rotation angle rather than the magnetic potential. In general, the vertex is the change of the Hamiltonian per unit rotation angle.

For a rotation through an angle \(\theta_i\) about a fixed transverse axis \(\mathbf u_i\),

\begin{equation}
\mathbf{e}_{i}\left( \theta_{i} \right) = \mathbf{e}_{i} + \theta_{i}\left( \mathbf{u}_{i} \times \mathbf{e}_{i} \right) + \frac{\theta_{i}^{2}}{2}\mathbf{u}_{i} \times \left( \mathbf{u}_{i} \times \mathbf{e}_{i} \right) + O\left( \theta_{i}^{3} \right)
\label{eq:orig009}
\end{equation}
Since \(\mathbf u_i\) is perpendicular to \(\mathbf e_i\), \(\mathbf u_i\)\(\times (\)\(\mathbf u_i\)\(\times\)\(\mathbf e_i\)\() = -\)\(\mathbf e_i\), and the corresponding local second-order perturbation is

\begin{equation}
V_{i}^{(2)} = \frac{b_{i}}{2}\theta_{i}^{2}\,\mathbf{e}_{i} \cdot \bm{\sigma}
\label{eq:orig010}
\end{equation}
The direct second-order term changes the on-site curvature but not the coupling between different sites.

For prescribed-field rotations, we evaluate the one-electron grand potential $\Omega^B$ at fixed chemical potential. Its second derivative with respect to the rotation angles defines $H^B$. The same distinction between field and moment rotations applies beyond local one-electron models, although nonlocal operators and correlated electrons introduce additional terms. The relaxed moment curvature $H^M$ instead refers to fixed directions of the integrated atomic moments $\mathbf M_i$. It does not prescribe the direction of the averaged exchange-correlation field or the single-site scattering potential. The field curvature is

\begin{equation}
H_{ij}^{B,ab} = \left. \ \frac{\partial^{2}\Omega^B}{\partial\theta_{i}^{a}\partial\theta_{j}^{b}} \right|_{0}
\label{eq:orig011}
\end{equation}
so that

\begin{equation}
\delta^{2}\Omega^B = \frac{1}{2}\sum_{ij}^{}{\sum_{ab}^{}{\theta_{i}^{a}H_{ij}^{B,ab}\theta_{j}^{b}}}
\label{eq:orig012}
\end{equation}
For two independently tilted directions, the second derivative of the electronic grand potential at fixed chemical potential is

\begin{equation}
\begin{aligned}
H_{ij}^{B,ab}={}&\sum_n f_n
 \left\langle n\left|\frac{\partial^2 \widehat H}
 {\partial\theta_i^a\partial\theta_j^b}\right|n\right\rangle\\
&+\frac12\sum_{nm}\frac{f_n-f_m}{\varepsilon_n-\varepsilon_m}
 \Bigl[\langle n|V_i^a|m\rangle\langle m|V_j^b|n\rangle\\
&\hspace{3em}+\langle n|V_j^b|m\rangle
 \langle m|V_i^a|n\rangle\Bigr].
\end{aligned}
\label{eq:orig013}
\end{equation}
Equation~\eqref{eq:orig013} contains both contributions to the second derivatives of the rotational energy: the direct second-order change in the Hamiltonian and the electronic response to two first-order vertices. Electronic transitions between occupied and unoccupied states generate the off-site interaction.

This form parallels earlier perturbative treatments \cite{ref1,ref2,ref3} and more recent exchange formulas \cite{ref13,ref16,ref18}. Because both transverse vertices are kept, we treat noncollinear and collinear states on the same footing, without assuming a Heisenberg model.

For distinct sites,

\begin{equation}
\frac{\partial^{2}\widehat{H}}{\partial\theta_{i}^{a}\partial\theta_{j}^{b}} = 0,\quad\quad i \ne j
\label{eq:orig014}
\end{equation}
and the second derivative for rotations at different sites is therefore

\begin{equation}
\begin{aligned}
H_{ij}^{B,ab}={}&\frac12\sum_{nm}
 \frac{f_n-f_m}{\varepsilon_n-\varepsilon_m}
 \Bigl[\langle n|V_i^a|m\rangle\langle m|V_j^b|n\rangle\\
&\hspace{3em}+\langle n|V_j^b|m\rangle
 \langle m|V_i^a|n\rangle\Bigr],\quad i\ne j.
\end{aligned}
\label{eq:orig015}
\end{equation}
The field pair curvature can also be calculated from the retarded one-electron Green function \(G^R(\varepsilon)=[\varepsilon+i0^+-H_{\mathrm{ref}}]^{-1}\), where \(H_{\mathrm{ref}}\) is the complete spin-dependent reference Hamiltonian, including its exchange splitting (not merely the spin-independent part). At zero temperature,

\begin{equation}
\begin{aligned}
H_{ij}^{B,ab} &=-\frac{1}{2\pi}\operatorname{Im}\int^{E_F}d\varepsilon\,\operatorname{Tr}\bigl[ V_i^aG_{ij}V_j^bG_{ji}\\ &\qquad +V_j^bG_{ji}V_i^aG_{ij}\bigr],\qquad i\ne j
\end{aligned}
\label{eq:orig016}
\end{equation}
Equation~\eqref{eq:orig016} represents the intersite response through two rotational vertices and the full reference Green function, including both local transverse directions in a noncollinear state. The two orderings give the same trace, so the symmetrized expression with prefactor \(1/(2\pi)\) is equivalent to the unsymmetrized expression with prefactor \(1/\pi\). The spectral and Green-function forms use the same static perturbation; in metals, degeneracies and Fermi-surface terms require the appropriate static limit. The integral is evaluated at zero temperature.

The relation to moment interactions is developed in Sec.~\ref{sec:pt_moment} and for ferromagnets in Sec.~\ref{sec:pt_fm}.

Local transverse axes are arbitrary; comparisons between exchange calculations require consistent rotation coordinates and pair-counting conventions.

\paragraph*{RPT and the LFT.}
For a prescribed rotation of the magnetic potential, ordinary perturbation theory gives $\partial\Omega^B/\partial\theta_i^a=\langle V_i^a\rangle$ at fixed chemical potential and the second-order energy curvature derived above. Neither result requires the LFT. The conventional magnetic LFT addresses a different question: when the first-order change of the self-consistent total energy can be obtained from a frozen-potential one-electron calculation. To identify the energy associated with the prescribed rotation, we consider the frozen-input Harris--Foulkes expression in LSDA.

For this purpose, we take locally rotated input spin densities and their exchange-correlation fields in LSDA, initially without SOC. The double-counting terms are unchanged along this prescribed rotation path. Let the input charge and spin densities be $n_0(\mathbf r)$ and $\mathbf m_{\rm in}(\mathbf r,\boldsymbol\theta)=\mathcal R_i(\boldsymbol\theta_i)\mathbf m_0(\mathbf r)$ in site region $i$, where $\mathcal R_i$ is a rigid spin rotation. The LSDA exchange field rotates with the input magnetization, $\mathbf B_{\rm xc}(\mathbf r,\boldsymbol\theta)=\mathcal R_i(\boldsymbol\theta_i)\mathbf B_{\rm xc}^0(\mathbf r)$, with $\mathbf B_{\rm xc}^0$ collinear with $\mathbf m_0$ and Hamiltonian magnetic term $-\mathbf B_{\rm xc}\cdot\bm\sigma$. The scalar effective potential, $n_0$, $|\mathbf m_0|$, and $|\mathbf B_{\rm xc}^0|$ remain fixed along this path.
For the same input density, the spin-polarized Harris--Foulkes energy \cite{Harris1985,FoulkesHaydock1989} has the form
\begin{equation}
\begin{split}
 E_{\rm HF}^B(\boldsymbol\theta)={}&E_{\rm band}(\boldsymbol\theta)-E_{\rm H}[n_0]
 +E_{\rm xc}[n_0,m_0]\\
 &-\int d\mathbf r\,n_0v_{\rm xc}^0
 +\int d\mathbf r\,\mathbf m_{\rm in}\cdot\mathbf B_{\rm xc}+E_{\rm ion}.
\end{split}
\label{eq:harris_foulkes_rot}
\end{equation}
Here $E_{\rm band}(\boldsymbol\theta)$ is the sum of occupied eigenvalues of the frozen-input Hamiltonian along the prescribed exchange-field rotational path; the displayed magnetic double-counting sign follows from the negative exchange-field term in that Hamiltonian. LSDA depends on the magnitude, rather than the direction, of the input magnetization. Thus $E_{\rm H}$, $E_{\rm xc}$, $v_{\rm xc}^0$, and $\mathbf m_{\rm in}\cdot\mathbf B_{\rm xc}=\mathbf m_0\cdot\mathbf B_{\rm xc}^0$ are independent of the rotation angles. Consequently,
\begin{equation}
 E_{\rm HF}^B(\boldsymbol\theta)-E_{\rm HF}^B(0)
 =E_{\rm band}(\boldsymbol\theta)-E_{\rm band}(0).
\label{eq:harris_foulkes_identity}
\end{equation}
Equation~\eqref{eq:harris_foulkes_identity} shows that the Harris--Foulkes and frozen-input band energies have identical first and second derivatives along this path. Thus RPT gives the prescribed-field Harris--Foulkes curvature when the same particle-number condition is used. This identity does not extend the first-order LFT to second order. 
Solovyev and Terakura \cite{refSolovyevTerakura1998} gave a related gauge-symmetry argument for small nonuniform rotations of the spin magnetization and the corresponding effective magnetic potential. Their result includes corrections from induced changes in charge density and magnetization magnitude. The order of those corrections must be assessed for the reference state and perturbation considered.
The choice of particle-number condition matters. At fixed electron number, the Hessian of this frozen-input energy is the fixed-$N$ band-energy Hessian. Equation~\eqref{eq:orig013} is instead the fixed-$\mu$ grand-potential Hessian. These coincide when the first-order particle-number response to the rotations vanishes. More generally, for a regular metallic response, the difference $H^B_{N,ab}-H^B_{\mu,ab}$ is $(\partial_{\theta_a}N)_\mu(\partial_{\theta_b}N)_\mu/(\partial_\mu N)_\theta$. Thus the two curvatures can be compared only at the same fixed particle number or chemical potential. For a collinear spin-rotation-invariant reference, the transverse first-order vertices give zero first-order charge response, and the two Hessians agree.
The rotated input spin density and self-consistent magnetization can differ already at first order in the rotation angles. Their difference can affect the second-order energy and hence its curvature. The moment curvature $H^M$ instead follows from the relaxed response to prescribed moment directions, including the required constraining fields; its relation to $H^B$ depends on that separate variational problem. The present frozen-input identity is not a second-order theorem for arbitrary constrained-moment rotations. The double-counting argument also survives the addition of a fixed SOC operator to the frozen-input Hamiltonian, provided the scalar input density and local magnetization magnitude remain unchanged and the SOC contribution is included consistently in the one-electron spectrum. SOC nevertheless breaks global spin-rotation invariance relative to the lattice; the nonrelativistic uniform-rotation zero-mode argument must not be applied to that case.

To compare the electronic rotational Hessian with a spin model, one can use the isotropic bilinear Heisenberg Hamiltonian
\begin{equation}
\mathcal H_{\mathrm H}=-\frac12\sum_{i\ne j}J_{ij}\,\mathbf e_i\cdot\mathbf e_j,\qquad |\mathbf e_i|=1.
\label{eq:heisenberg_definition}
\end{equation}
Here $J_{ij}=J_{ji}$ multiplies unit moment directions rather than full moment amplitudes, $J_{ij}>0$ favors parallel alignment, and the half-sum avoids double counting. This model predicts a particular mixed second derivative under two local rotations. For a general bilinear exchange tensor $\mathcal J_{ij}^{\alpha\beta}$ in the same half-sum convention, the off-site spin-model Hessian is its restriction to rotations perpendicular to the reference moments. Its comparison with $H^{B,ab}_{ij}$ or $H^{M,ab}_{ij}$ requires the corresponding choice of rotational variables:

\begin{equation}
H_{ij}^{\mathrm{spin},ab} =  - \ \sum_{\alpha\beta}^{}{\ t_{i\alpha}^{a}\ \mathcal{J}_{ij}^{\alpha\beta}\ t_{j\beta}^{b}}.
\label{eq:orig017}
\end{equation}
Equation~\eqref{eq:orig017} follows from the spin model, not from electronic perturbation theory. The isotropic component of a specified bilinear exchange tensor is $J_{ij}^{\mathrm{iso}}=\frac13\sum_{\alpha}\mathcal J_{ij}^{\alpha\alpha}$; its remaining symmetric traceless and antisymmetric components describe anisotropic bilinear couplings. For the isotropic Heisenberg model, $\mathcal J_{ij}^{\alpha\beta}=J_{ij}\delta_{\alpha\beta}$, and Eq.~\eqref{eq:orig017} reduces to

\begin{equation}
H_{ij}^{\mathrm{spin},ab} =  - \ J_{ij}\ P_{ij}^{ab},\ \ \ \ \ P_{ij}^{ab} = \mathbf{t}_{i}^{a}\  \cdot \ \mathbf{t}_{j}^{b}.
\label{eq:orig018}
\end{equation}

\paragraph*{Mapping the electronic Hessian onto spin interactions.}
RPT gives energy derivatives without assuming a spin Hamiltonian. For a given magnetic state, transverse rotations sample only four combinations of the nine exchange-tensor components. Different exchange tensors can therefore give the same energy change for small transverse rotations. For the isotropic Heisenberg form in Eq.~\eqref{eq:orig018}, however, all transverse components must satisfy $H_{ij}^{\mathrm{spin},ab}=-J_{ij}P_{ij}^{ab}$ with the same $J_{ij}$, including components for which $P_{ij}^{ab}=0$. Section~III develops this test for noncollinear states and identifies the part of the electronic Hessian that cannot be represented by isotropic pair exchange.

The curvature at one magnetic configuration does not uniquely determine a spin Hamiltonian. Different magnetic interactions can give the same curvature. Without SOC, a bilinear pair Hamiltonian with fixed, configuration-independent coefficients is restricted by global spin-rotation symmetry to the isotropic Heisenberg form. A single spin Hamiltonian must reproduce the curvatures at different magnetic configurations. The comparison must use the same electronic functional, magnetic variables, and relaxation conditions. It must also account for on-site curvatures and torques. At constrained nonstationary configurations, the geometric terms must also be included. Agreement beyond small rotations also requires the correct energy at finite rotation angles. Section~IV applies the Hessian comparison to spin spirals.

Independently of any spin Hamiltonian, global spin-rotation symmetry constrains the electronic energy derivatives. For a common spin rotation of the complete one-electron Hamiltonian, $\hat H(\lambda)=U(\lambda)\hat H_0U^\dagger(\lambda)$ with $U(\lambda)=e^{-i\lambda S}$, the perturbation is $V=-i[S,\hat H_0]$. Since the equilibrium density matrix is a function of $\hat H_0$, its first-order energy variation vanishes: $\partial_\lambda\Omega^B=\operatorname{Tr}(\rho V)=0$. This identity expresses the vanishing total torque for a global rotation; it does not require the LFT and does not imply that individual local torques vanish. The corresponding second-order constraint is derived below.

The local second-order contribution is required by global spin-rotation invariance. In the absence of SOC, the same infinitesimal rotation \(\delta\boldsymbol{\Theta}\) of all spins about a fixed laboratory axis changes each reference direction according to

\begin{equation}
\delta\mathbf{e}_{i} = \delta\mathbf{\Theta}\  \times \ \mathbf{e}_{i}
\label{eq:orig019}
\end{equation}
The vector \(\delta\boldsymbol{\Theta}\) is common in the laboratory frame, but its transverse components \(\Theta_{i}^{a}\) in the local tangent basis depend on the site. The deviations can therefore be written

\begin{equation}
\delta\mathbf{e}_{i} = \sum_{a}^{}{\Theta_{i}^{a}\mathbf{t}_{i}^{a}}
\label{eq:orig020}
\end{equation}
At a configuration with zero magnetic torque, these uniform rotations leave the energy unchanged to second order, so

\begin{equation}
\sum_{jb}^{}{H_{ij}^{ab}\Theta_{j}^{b}} = 0
\label{eq:orig021}
\end{equation}
For a general configuration, write $g_A=\partial E/\partial\theta_A$ and $H_{BA}=\partial^2E/(\partial\theta_B\partial\theta_A)$, with $A=(i,a)$, and let $v_A^{(\mu)}(\boldsymbol\theta)$ denote the angular change generated by a global laboratory rotation about axis $\mu$. Rotational invariance gives $\sum_Ag_Av_A^{(\mu)}=0$; differentiation yields
\begin{equation}
 \sum_A H_{BA}v_A^{(\mu)}
 =-\sum_A g_A\frac{\partial v_A^{(\mu)}}{\partial\theta_B}.
 \label{eq:stage2Ward}
\end{equation}
The rotation components $v_A^{(\mu)}$ depend on the local transverse frames. A global rotation leaves the energy unchanged for any reference configuration, but its angular displacement is a zero mode of the coordinate Hessian only at a stationary configuration of the same energy functional ($g_A=0$). At such a configuration the right-hand side vanishes and Eq.~\eqref{eq:orig021} follows; otherwise the energy gradient contributes. For a collinear state with a common transverse basis, the stationary identity reduces to the usual row-sum rule. Here $g_A$ is the gradient with respect to the chosen angular coordinates; its negative is the generalized torque conjugate to those coordinates, not necessarily a fixed laboratory-frame torque component.

The global-rotation identity can also be verified directly within RPT using the unitary transformation introduced above. Its second-order vertex is $W=-[S,[S,\hat H_0]]$. In the eigenbasis of $\hat H_0$, the two terms in the second-order energy derivative of an eigenstate obey
\begin{equation}
\begin{split}
\langle n|W|n\rangle
 &=-2\sum_m(\varepsilon_n-\varepsilon_m)|S_{nm}|^2,\\
2\sum_{m\ne n}\frac{|V_{nm}|^2}{\varepsilon_n-\varepsilon_m}
 &=+2\sum_{m\ne n}(\varepsilon_n-\varepsilon_m)|S_{nm}|^2.
\end{split}
\label{eq:pt_global_cancellation}
\end{equation}
With a complete eigenstate basis, the direct second-order vertex and the response to the first-order vertex cancel for each eigenstate and hence in the occupied-state sum. This is a direct check that RPT respects global spin-rotation invariance: retaining only the two-vertex response would violate the identity. For the prescribed rigid-input path in local LSDA without SOC, the Harris--Foulkes identity, Eq.~\eqref{eq:harris_foulkes_identity}, gives the same uniform-rotation curvature. This cancellation applies to uniform rotations; it neither requires vanishing curvature for nonuniform rotations nor establishes equality between the prescribed-field and constrained-moment Hessians. For nonlocal magnetic operators the rotation must include all induced vertices and contact terms; relaxed or correlated response problems require their own consistent electronic variations.

The indices \(a,b\) label the two transverse directions at each site. The individual matrix elements depend on the choice of these directions. Under a local change of transverse coordinates,

\begin{equation}
{\theta'}_{i}^{a} = \sum_{c}^{}{\left( \mathbf{R}_{i} \right)_{ac}\,\theta_{i}^{c}}
\label{eq:orig022}
\end{equation}
The first-order vertices transform as $V'_{ia}=\sum_c(R_i)_{ac}V_{ic}$, and the direct second-order vertex transforms with one factor of $R$ per angular index. Thus both terms in Eq.~\eqref{eq:orig013} give the transformation rule

\begin{equation}
{H'}_{ij} = \mathbf{R}_{i}H_{ij}\mathbf{R}_{j}^{T}
\label{eq:orig023}
\end{equation}
The rotational energy in Eq.~\eqref{eq:orig012} does not depend on how the transverse axes are chosen. Changing these axes is not the same as physically rotating the spins.

\subsection{Rotations of the magnetization}
\label{sec:pt_moment}

We now consider rotations of the magnetic moments rather than prescribed exchange fields. The moment directions are fixed by constraints while the electronic state responds. For a perturbation \(H(\lambda)=H_0+\lambda V+\lambda^2 W/2+\cdots\), ordinary nondegenerate many-electron perturbation theory gives the first-order wave-function correction \( |0^{(1)}\rangle=\sum_{n\ne0}|n\rangle\langle n|V|0\rangle/(E_0-E_n)\). Its magnetization response is

\begin{equation}
M_{\alpha}^{(1)}\left( \mathbf{r} \right) = 2\operatorname{Re}\sum_{n\neq 0}^{}\frac{\langle 0\left| {\widehat{M}}_{\alpha}\left( \mathbf{r} \right) \right|n\rangle\langle n|V|0\rangle}{E_{0} - E_{n}}
\label{eq:orig024}
\end{equation}
A prescribed change in magnetization thus requires a corresponding constraining field, whose variation is determined by the electronic response. We use the same rotation angles as above, but here \(\mathbf e_i\) denotes the direction of the integrated atomic moment, not necessarily that of the prescribed exchange field. The fixed-length moment path is \(\mathbf M_i(\theta)=M_i\mathbf e_i(\theta)\). Let \(\mathbf{t}_{i}^{a}\) and \(\mathbf s_{i,ab}\) be the first and second derivatives of \(\mathbf e_i\) in the same rotational parametrization.

\begin{equation}
\frac{\partial\mathbf{M}_{i}}{\partial\theta_{j}^{a}} = \delta_{ij}M_{i}\mathbf{t}_{i}^{a},\quad\quad\frac{\partial^{2}\mathbf{M}_{i}}{\partial\theta_{j}^{a}\partial\theta_{k}^{b}} = \delta_{ij}\delta_{ik}M_{i}\mathbf{s}_{i,ab}
\label{eq:orig025}
\end{equation}
Edwards \cite{ref14} related the electronic interaction function to the response of the constraining fields required to maintain a prescribed spin-density configuration. The transverse fields balance the electronic torques associated with changes in magnetization. A Heisenberg-like exchange follows only after an additional approximation for the spin dependence of the electronic interaction function. Here we constrain integrated atomic moments while allowing the internal magnetization distribution to relax. We introduce fields \(\mathbf{B}_{i}^{c}\) conjugate to these moments, with \(V^c=-\sum_i\mathbf B_i^c\cdot\widehat{\mathbf M}_i\). To calculate the response, the reference constraining fields are included in the auxiliary Hamiltonian and their variations are treated as perturbations:

\begin{equation}
\delta M_{i\alpha} = \sum_{j\beta}^{}{X_{i\alpha,j\beta}\delta B_{j\beta}^{c}},\quad\quad\delta\mathbf{B}^{c} = X^{- 1}\delta\mathbf{M}
\label{eq:orig026}
\end{equation}
Here \(X\) is the static magnetic susceptibility of the specified local moments to their conjugate fields, with all other electronic degrees of freedom allowed to respond. Equation~\eqref{eq:orig024} gives the corresponding many-electron response when the perturbation is the variation of the constraining field. The independent-particle susceptibility omits the induced changes in the electronic potential.

We denote the ground-state energy of the auxiliary Hamiltonian, including its field coupling, by \(\mathcal E(\mathbf B^c)\). The Hellmann--Feynman theorem gives \(\partial\mathcal E/\partial B_{i\alpha}^c=-M_{i\alpha}\). Because the auxiliary field enters the Hamiltonian linearly, second-order perturbation theory gives Eq.~\eqref{eq:orig027}. The quadratic energy change itself contains an additional factor of 1/2: \(\Delta\mathcal E^{(2)}=\delta^2\mathcal E/2\).

\begin{equation}
\delta^{2}\mathcal E = -\sum_{i\alpha,j\beta}^{}{\delta B_{i\alpha}^{c}X_{i\alpha,j\beta}\delta B_{j\beta}^{c}}
\label{eq:orig027}
\end{equation}
Here \(\mathcal E\) includes the auxiliary-field coupling, whereas \(E\) is the electronic energy of the state selected by the constraints, with that coupling subtracted. The Legendre transform of \(\mathcal E\) and the first differential of \(E\) are

\begin{equation}
E\left(\mathbf M\right)=\mathcal E\left(\mathbf B^c\right) + \sum_{i}^{}{\mathbf{B}_{i}^{c} \cdot \mathbf{M}_{i}},\quad\quad dE = \sum_{i}^{}{\mathbf{B}_{i}^{c} \cdot d\mathbf{M}_{i}}
\label{eq:orig028}
\end{equation}
Where the moment--field response is locally invertible, the second derivative of the constrained total energy with respect to the moments is the inverse susceptibility. When global spin-rotation symmetry makes the response singular, the inverse is taken after excluding the zero-energy rotations. Equivalently, a small symmetry-breaking field can be introduced and removed after the inversion. The resulting second derivative is

\begin{equation}
\frac{\partial^{2}E}{\partial M_{i\alpha}\partial M_{j\beta}} = \left( X^{- 1} \right)_{i\alpha,j\beta}
\label{eq:orig029}
\end{equation}
Differentiating with respect to the rotation angles gives \(H^M\), the Hessian of the relaxed constrained density-functional total energy:

\begin{equation}
H_{ia,jb}^{M} = M_{i}M_{j}\mathbf{t}_{i}^{a} \cdot \left( X^{- 1} \right)_{ij}\mathbf{t}_{j}^{b} + \delta_{ij}M_{i}\mathbf{B}_{i}^{c} \cdot \mathbf{s}_{i,ab}
\label{eq:orig030}
\end{equation}
In Eq.~\eqref{eq:orig030}, the first term comes from the inverse response; the second arises because the moment direction changes nonlinearly with the rotation angle. At a stationary reference with vanishing conjugate fields, the geometric term disappears. A longitudinal constraining field, however, can contribute to the angular curvature even when the rotational torque vanishes. In contrast, Eq.~\eqref{eq:orig013} differentiates the grand potential of a prescribed rotating one-electron Hamiltonian. The two curvatures describe different electronic responses and generally differ.

The effect of internal electronic relaxation can be illustrated by two magnetic angles \(\theta_{1}\) and \(\theta_{2}\) coupled to an internal transverse electronic degree of freedom \(x\). Near a stationary reference, one can consider the quadratic energy:

\begin{equation}
\begin{aligned}\Delta E={}&(A/2)(\theta_1-\theta_2)^2\\&+g(\theta_1-\theta_2)x+(C/2)x^2,\quad C>0.\end{aligned}
\label{eq:relax_example_energy}
\end{equation}

Holding \(x\) fixed gives the relative-angle curvature \(A\). Minimizing over \(x\) gives the electronic displacement and the reduced curvature:

\begin{equation}
x_{\mathrm{rel}} =  - (g/C)(\theta_{1}\  - \ \theta_{2}),\ \ \ \ \ H_{\mathrm{rel}} = A\  - \ g^{2}/C.
\label{eq:relax_example_result}
\end{equation}

Electronic relaxation softens the relative-rotation mode but leaves uniform rotation as a zero mode. For \(A>0\) and \(C>0\), stability against relative rotations and internal-coordinate fluctuations requires \(AC>g^2\). More generally, let $\theta$ describe the moment rotations and $x$ the internal electronic degrees of freedom. When the electrons are allowed to relax at each moment direction, their response reduces the magnetic energy curvature to

\begin{equation}
H_{\mathrm{rel}} = A\  - \ GC^{- 1}G^{T}.
\label{eq:relax_schur}
\end{equation}

The internal electronic degrees of freedom eliminated in Eq.~\eqref{eq:relax_schur} may include relative transverse responses of electronic states with different orbital character. For example, the \(t_{2g}\), \(e_g\), and more itinerant \(sp\) contributions to the magnetization of bcc Fe need not rotate rigidly under a nonuniform magnetic perturbation. A conventional Heisenberg Hamiltonian contains one magnetic direction per atomic site, whereas the susceptibility of the total atomic moments includes these internal responses. Thus, internal electronic relaxation modifies the moment-rotation curvature, although this correction cannot generally be identified with \(H^M-H^B\).

Ref. \cite{ref22} applied the inverse static susceptibility directly to first-principles exchange calculations. Equations~\eqref{eq:orig029} and \eqref{eq:orig030} link it to the moment-component and angular energy derivatives of the constrained moments, respectively.

Without SOC, simultaneous global rotation of all moments leaves the energy unchanged. At a nonstationary constrained configuration, rotational invariance yields a torque identity whose derivative also contains first energy derivatives; the angular Hessian need not have a global-rotation zero mode. At an unconstrained stationary magnetic state without SOC or external symmetry-breaking fields, uniform rotation is a zero mode of the angular curvature. The same zero mode appears in the inverse transverse susceptibility when the response preserves spin-rotation symmetry and the chosen moment variables retain the collective rotation.

\subsection{Magnetic torques and constraining fields}
\label{sec:pt_torque}

The first-order rotational vertex \(V_{i}^{a}\) introduced above gives the field-rotation torque directly from electronic perturbation theory:

\begin{equation}
\tau_{ia}^{B} =  - \frac{\partial E^{B}}{\partial\theta_{i}^{a}} =  - \langle V_{i}^{a}\rangle
\label{eq:orig031}
\end{equation}
Equation~\eqref{eq:orig031} gives the first derivative of the field energy. It need not equal the torque derived from the fully relaxed moment energy.

For the magnetization, the constraining field introduced in Eq.~\eqref{eq:orig026} is conjugate to the constrained local moment. The Legendre transform in Eq.~\eqref{eq:orig028} gives \(dE=\sum_i\mathbf B_i^c\cdot d\mathbf M_i\). Under an infinitesimal rigid rotation \(\delta\mathbf M_i=\delta\boldsymbol\theta_i\times\mathbf M_i\), defining \(\delta E=-\sum_i\boldsymbol\tau_i^M\cdot\delta\boldsymbol\theta_i\) yields

\begin{equation}
\tau_{i}^{M} =  - M_{i}\  \times \ \mathbf{B}_{i}^{c}\ ;\ \ \ \ \ \tau_{ia}^{M} =  - M_{i}\ \mathbf{B}_{i}^{c}\  \cdot \ \mathbf{t}_{i}^{a}
\label{eq:orig032}
\end{equation}
\begin{equation}
H_{ia,jb}^{M} =  - \ \frac{\partial\tau_{ia}^{M}}{\partial\theta_{j}^{b}}
\label{eq:orig033}
\end{equation}
The constraining field balances the torque acting on the moment; only its transverse component contributes to the first-order torque. Here \(\tau_{ia}^{M}=-\partial E/\partial\theta_i^a\) is the generalized torque conjugate to the rotation coordinate \(\theta_i^a\), not a fixed laboratory component. At an unconstrained stationary configuration, the rotational torques vanish.

In stationary constrained noncollinear LSDA without SOC, the local exchange-correlation field produces no torque because it is parallel to the magnetization. The integrated spin-current flux is balanced by the torque of the constraining field (and by any other explicitly applied field torques). For an unconstrained stationary state without external fields, the net spin-current flux through each chosen atomic region instead vanishes. Variations of this torque balance determine the restoring torques for moment rotations, while their dynamics require the frequency-dependent electronic response.

The adiabatic approximation in spin dynamics \cite{ref23,ref24} separates the slow collective rotation of the magnetic moments from the electronic response. In a two-time response function, $t_2$ is the time of the perturbation and $t_1$ is the time at which the induced response is evaluated. The center time $\bar t=(t_1+t_2)/2$ labels the evolving magnetic configuration, while the relative time $\Delta t=t_1-t_2$ describes the electronic response interval. Writing $\mathbf M_i(\bar t)=M_i(\bar t)\mathbf e_i(\bar t)$, with $|\mathbf e_i(\bar t)|=1$, the separation requires the direction to change little over relative times that contribute appreciably to the response:
\begin{equation}
\begin{aligned}
\mathbf e_i(\bar t+\Delta t/2)
 &=\mathbf e_i(\bar t)
   +\frac{\Delta t}{2}\dot{\mathbf e}_i(\bar t)
   +O(\Delta t^2),\\
|\Delta t\,\dot{\mathbf e}_i(\bar t)|&\ll 1.
\end{aligned}
\label{eq:adiabaticTimescale}
\end{equation}
If longitudinal fluctuations can also be neglected, the additional fixed-magnitude approximation gives $M_i(\bar t)\simeq M_i^{(0)}$ and $\mathbf M_i(\bar t)\simeq M_i^{(0)}\mathbf e_i(\bar t)$; this approximation does not follow from the separation of electronic and magnetic time scales. At leading adiabatic order, one can solve stationary density-functional Pauli equations for successive configurations stabilized by constraining fields. Equation~\eqref{eq:orig032} then gives the instantaneous static torque, whereas precession requires a dynamical coefficient: the static inverse susceptibility determines the moment-rotation energy derivatives, and its frequency-linear term determines the leading low-frequency motion. This coefficient need not equal that of the prescribed-field response in Sec.~\ref{sec:pt_time}.

Near a stationary state, small oscillations can alternatively be obtained directly from the dynamical transverse susceptibility, without prescribing successive moment directions.

\subsection{Time-dependent RPT}
\label{sec:pt_time}

Extending the time-dependent treatments of Refs. \cite{ref9,ref10,ref19}, we now allow the exchange-field rotations to oscillate in time. Their first-order perturbation is

\begin{equation}
V^{(1)}(t) = \sum_{ia}^{}{\theta_{i}^{a}(t)V_{i}^{a}},\quad\quad\theta_{i}^{a}(t) = \theta_{i}^{a}(\omega)e^{- i\omega t}
\label{eq:orig034}
\end{equation}
The response then follows from the same first-order vertices, with the full-Hessian normalization of Eq.~\eqref{eq:orig013}:

\begin{equation}
\mathcal{K}_{ia,jb}(\omega) = \langle W_{ia,jb}\rangle + \Pi_{ia,jb}(\omega)
\label{eq:orig035}
\end{equation}
\begin{equation}
\Pi_{ia,jb}(\omega) = \sum_{nm}^{}\frac{\left( f_{n} - f_{m} \right)\left( V_{i}^{a} \right)_{nm}\left( V_{j}^{b} \right)_{mn}}{\varepsilon_{n} - \varepsilon_{m} + \hbar\omega + i0^{+}}
\label{eq:orig036}
\end{equation}
The occupations and state normalization are the same as in the static calculation. At finite frequency, the two vertex orderings differ; the regular zero-frequency limit recovers the static Hessian.

\begin{equation}
\mathcal{K}_{ia,jb}(0) = H_{ia,jb}\quad\text{(strictly static limit)}
\label{eq:orig037}
\end{equation}
Equation~\eqref{eq:orig037} gives the field Hessian in this limit. In a metal, the zero-frequency and zero-wave-vector limits can differ, and a pole of the field response need not be a collective spin-wave mode.

When the electronic response is regular at low frequency, it admits the expansion

\begin{equation}
\mathcal{K}\left( \mathbf{q},\omega \right) = H\left( \mathbf{q} \right) + \hbar\omega\Lambda\left( \mathbf{q} \right) + O\bigl((\hbar\omega)^{2}\bigr)
\label{eq:orig038}
\end{equation}
Electron--hole excitations can damp spin waves and spoil a regular low-frequency expansion. Collective modes must then be found from the full self-consistent transverse susceptibility.

\subsection{Frequency-dependent intersite interactions}

We now use the retarded kernel to describe the frequency dependence of intersite rotational response. For $i\ne j$, we define an \emph{exchange-like} transverse tensor using the angular-curvature and pair-counting convention of Eq.~\eqref{eq:heisenberg_definition}:
\begin{equation}
\mathcal J_{ij}^{ab}(\omega)\equiv-\mathcal{K}_{ia,jb}^{R}(\omega),\qquad i\ne j.
\label{eq:dynJ_tensor}
\end{equation}
The superscript $R$ denotes retarded response. For a collinear reference with consistently chosen local transverse axes, if the static field curvature admits an isotropic bilinear representation, then $\mathcal J_{ij}^{ab}(0)=J_{ij}^{B}\delta_{ab}$. Here $J_{ij}^{B}$ is the prescribed-field exchange-like coefficient and is not, in general, the constrained-moment coefficient $J_{ij}^{M}$. Outside this scalar limit, the response is a tensor. Its frequency dependence does not define a frequency-dependent Heisenberg Hamiltonian.

Writing $z=\hbar\omega$, a regular low-frequency limit of Eq.~\eqref{eq:orig036} yields
\begin{equation}
\begin{aligned}
\mathcal{K}_{ia,jb}^{R}(z)&=H_{ia,jb}+z\Lambda_{ia,jb}^{R}+o(z),\\
\Lambda_{ia,jb}^{R}&=\left.\frac{\partial\Pi_{ia,jb}^{R}(z)}{\partial z}\right|_{z=0}\\
&=-\lim_{z\to0}\sum_{nm}\frac{(f_n-f_m)(V_i^a)_{nm}(V_j^b)_{mn}}{(\varepsilon_n-\varepsilon_m+z+i0^+)^2}.
\end{aligned}
\label{eq:dynJ_slope}
\end{equation}
For prescribed rotations of a static one-electron Hamiltonian, the second-order vertex $W_{ia,jb}$ has no explicit frequency dependence. When the frequency derivative of the response is well defined and dissipation is absent, the additional energy denominator in Eq.~\eqref{eq:dynJ_slope} gives a spatial dependence different from that of the static exchange. In metals, however, the low-frequency retarded response may be nonanalytic, and the zero-frequency limit cannot generally be interchanged with the electronic-state summation.

For the collinear case considered here, the transverse response takes the form
\begin{equation}
\begin{aligned}
\mathcal{K}_{ij}^{R}(z)&=-J_{ij}^{B}I+iz\,g_{ij}\epsilon+iz\,\eta_{ij}I+o(z),\quad i\ne j,\\
\epsilon&=\begin{pmatrix}0&1\\-1&0\end{pmatrix}.
\end{aligned}
\label{eq:dynJ_cartesian}
\end{equation}
The coefficient $g_{ij}$ describes precession, while $\eta_{ij}$ describes the frequency-linear dissipative response when this expansion exists. These terms have different physical meanings even though both contain $i\omega$ in this representation. At positive frequency, an equilibrium system cannot supply net energy to an applied perturbation. Thus the full absorption matrix, rather than any individual off-site element, must satisfy $-\operatorname{Im}_H \mathcal{K}^R(\omega)\succeq0$ for $\omega>0$ in the present sign convention. For the low-frequency expansion this gives $-\eta\succeq0$ for the complete matrix, although an individual off-site $\eta_{ij}$ can have either sign.

To fix the circular labels, we take $e_\pm=(1,\pm i)^T/\sqrt2$, so that $\epsilon e_\pm=\pm i e_\pm$. The eigenvalues of Eq.~\eqref{eq:dynJ_cartesian} and the corresponding exchange-like projections are then
\begin{equation}
\begin{aligned}
\mathcal{K}_{ij,\pm}^{R}(z)&=-J_{ij}^{B}\mp z g_{ij}+iz\eta_{ij}+o(z),\\
J_{ij,\pm}(\omega)&\equiv-\mathcal{K}_{ij,\pm}^{R}(\omega)\\
&=J_{ij}^{B}\pm\hbar\omega\,g_{ij}-i\hbar\omega\,\eta_{ij}+o(\omega).
\end{aligned}
\label{eq:dynJ_circular}
\end{equation}
These $\pm$ labels specify the circular polarization vectors, not independently the positive-frequency magnon branch; the latter must be identified using the pole and Fourier conventions of the dynamical response.

An elementary test of the precessional coefficient is the one-electron Hamiltonian $H=-b\sigma_z$ ($b>0$), with its lower spin state occupied. For $V^x=-b\sigma_x$ and $V^y=-b\sigma_y$, the contact term $\langle W_{xx}\rangle=b$ cancels the zero-frequency transverse bubble $\Pi_{xx}^R(0)=-b$, while $\partial_z\Pi_{xy}^R(0)=-i/2$. Hence $\mathcal{K}_{xx}^R(0)=0$ and $\Lambda_{xy}^R=-i/2$, with no low-frequency absorption below the spin-flip gap. This confirms the cancellation of the static terms and the correct sign of the precessional contribution to the dynamical susceptibility.

The uniform spin-precession response checks the sign and normalization of the prescribed-field frequency-linear kernel in Eq.~\eqref{eq:orig036}, under the local static-splitting assumptions below. For a spin-rotation-invariant collinear reference with local static magnetic potential $-b\sigma_z$, the complete uniform vertices are $V^x=-b\sigma_x$ and $V^y=-b\sigma_y$, while $H^\downarrow-H^\uparrow=2b$. Between spin-up and spin-down eigenstates,
\begin{equation}
\langle n\uparrow|b|m\downarrow\rangle=\frac{\varepsilon_m^\downarrow-\varepsilon_n^\uparrow}{2}\langle n\uparrow|m\downarrow\rangle.
\label{eq:dynJ_uniform_vertex}
\end{equation}
The two spin-flip orderings in Eq.~\eqref{eq:orig036}, followed by completeness of both spin eigenstate sets, give
\begin{equation}
\begin{aligned}
\sum_{ij}\Lambda_{ix,jy}^{R}&=-\frac{i}{2}(N_\uparrow-N_\downarrow)=-\frac{iM}{2},\\
\sum_{ij}\Lambda_{iy,jx}^{R}&=+\frac{iM}{2},\qquad
\sum_{ij}\Lambda_{ix,jx}^{R}=0.
\end{aligned}
\label{eq:dynJ_uniform_sum}
\end{equation}
Thus, with $g_{ij}=\Lambda_{ix,jy}^{R}/i$ as in Eq.~\eqref{eq:dynJ_cartesian}, the full site sum satisfies $\sum_{ij}g_{ij}=-M/2$. For a translationally invariant ferromagnet with one magnetic site per cell, this becomes
\begin{equation}
g_{\mathrm{on}}+\sum_{n\ge1}z_n g_n=-\frac{m}{2},
\label{eq:dynJ_shell_sum}
\end{equation}
where $z_n$ is the number of sites in coordination shell $n$, and $m$ is the spin polarization per cell. The sum includes the on-site coefficient, although the exchange-like definition in Eq.~\eqref{eq:dynJ_tensor} applies only off site. It fixes the total precessional normalization, not any individual shell coefficient. In particular, $g_n$ need not be proportional to the corresponding static $J_n$.

The preceding sum rule concerns the prescribed-field kernel for a local, static splitting. The physical uniform response has an independent normalization. For a spin-rotation-invariant Hamiltonian, $[H,S^\pm]=0$ and $[S^+,S^-]=2S^z$ give, for the retarded total-spin response per cell,
\begin{equation}
\chi_{+-}^R(\mathbf 0,z)=\frac{2S}{z+i0^+},\qquad z=\hbar\omega,
\label{eq:physical_spin_commutator}
\end{equation}
where $S=\langle S^z_{\mathrm{tot}}\rangle/N$ in units of $\hbar$, and $0^+$ has energy units. For a transverse source $-\mathbf f\cdot\mathbf e=-\tfrac12(f^-e^++f^+e^-)$ with $e^+=S^+/S$, the response to $f^+$ is $\delta e^+=\chi_{+-}^R f^+/(2S^2)=f^+/(Sz)$. Thus the inverse response in the unit-vector Hessian convention has the frequency term $-Sz$ on the positive-frequency branch (with the retarded sign convention used here). This identity applies also to conserving correlated responses; a regular, isolated Goldstone mode and a consistent moment projection are required to extend it to small nonzero wave vector. Nonlocal rotational vertices and correlated self-energy responses must still be included in the respective electronic kernels.

A collective-mode frequency depends on both the static restoring curvature and the precessional response. Consequently, mapping a static energy onto a Heisenberg Hamiltonian and fitting a Heisenberg spin-wave dispersion need not give the same pair interactions. The former matches a specified energy curvature; the latter also assumes a spin length and a dynamical law. The coefficients agree when a single Heisenberg model, with a consistent spin normalization, describes both quantities. Retardation, damping, and additional electronic modes can prevent such a description. Fitted parameters also depend on the wave-vector range and assumed interaction range. The electronic projections $J_{ij,\pm}(\omega)$ introduced above are not themselves dispersion-fitted Heisenberg parameters.

\section{MAGNETIC INTERACTIONS FROM RPT}
\label{sec:pt_interactions}

\subsection{Microscopic noncollinear pair interaction}
\label{sec:pt_noncollinear}

We calculate the pair energy curvature of a noncollinear magnetic state using the rotational vertices of Sec.~\ref{sec:pt_repr}. The Green functions are evaluated for the chosen magnetic configuration.

For distinct sites, the field pair curvature is given by Eq.~\eqref{eq:orig016}. The same result can be written as
\begin{equation}
\begin{aligned}
H_{ij}^{B,ab}={}&-\frac{1}{\pi}\operatorname{Im}\int^{E_F}d\varepsilon\,
\operatorname{Tr}\left[V_i^aG_{ij}(\varepsilon)V_j^bG_{ji}(\varepsilon)\right],\\
&\hspace{2em}i\ne j.
\end{aligned}
\label{eq:orig039}
\end{equation}
The on-site second-order vertex enters the full Hessian through Eq.~\eqref{eq:orig013}, but not the distinct-site pair derivative.

Comparing Eq.~\eqref{eq:orig039} with the Heisenberg Hessian in Eq.~\eqref{eq:orig018}, each transverse channel gives an estimate of the isotropic exchange coefficient:

\begin{equation}
J_{ij}^{B,(ab)} = \frac{1}{\pi}\frac{\operatorname{Im}\int_{}^{E_{F}}{d\varepsilon\, \operatorname{Tr}\left\lbrack V_{i}^{a}G_{ij}V_{j}^{b}G_{ji} \right\rbrack}}{\mathbf{t}_{i}^{a} \cdot \mathbf{t}_{j}^{b}}
\label{eq:orig040}
\end{equation}
Equation~\eqref{eq:orig040} applies only when the geometrical denominator is nonzero. A single Heisenberg exchange describes the pair curvature only if all transverse components satisfy Eq.~\eqref{eq:orig018} with the same coefficient, including those with vanishing geometrical factors.

When the calculated Hessian does not have the Heisenberg form, we can still find the isotropic exchange coefficient that best reproduces its transverse components. Minimizing $\sum_{ab}(H_{ij}^{ab}+JP_{ij}^{ab})^2$ gives
\begin{equation}
J_{ij}^{\mathrm{proj}}=-\frac{\sum_{ab}H_{ij}^{ab}P_{ij}^{ab}}
{1+(\mathbf e_i\cdot\mathbf e_j)^2},
\label{eq:hessian_to_j_projection}
\end{equation}
since $\sum_{ab}(P_{ij}^{ab})^2=1+(\mathbf e_i\cdot\mathbf e_j)^2$.
The result does not depend on the choice of local transverse axes, even when some geometric factors vanish. It can be applied separately to $H^B$ and $H^M$, using the corresponding energy curvature and consistent rotation and pair-counting conventions.

For illustration, let $c=\cos\phi$ for two reference moments
separated by angle $\phi$. In tangent frames where the geometrical
matrix is diagonal, consider a \emph{general} calculated pair contribution:
\begin{equation}
P_{ij}=\begin{pmatrix}c&0\\0&1\end{pmatrix},\qquad
H_{ij}=\begin{pmatrix}h_{11}&h_{12}\\h_{21}&h_{22}\end{pmatrix}.
\label{eq:spin_hessian_matrix}
\end{equation}
Equation~\eqref{eq:hessian_to_j_projection} gives
\begin{equation}
J_{ij}^{\mathrm{proj}}=-\frac{h_{11}c+h_{22}}{1+c^2}.
\label{eq:hessian_to_j_example}
\end{equation}
The difference between the calculated pair curvature and its Heisenberg contribution is
\begin{equation}
\begin{aligned}
\Delta H_{ij}&=H_{ij}+J_{ij}^{\mathrm{proj}}P_{ij}\\
&=\begin{pmatrix}
d_{ij}&h_{12}\\
h_{21}&-c\,d_{ij}
\end{pmatrix},
\qquad
d_{ij}=\frac{h_{11}-ch_{22}}{1+c^2}.
\end{aligned}
\label{eq:spin_hessian_transform}
\end{equation}
The pair curvature reduces to the Heisenberg form when $h_{11}=ch_{22}$ and $h_{12}=h_{21}=0$. The off-diagonal components do not enter the projected coefficient but still contribute to the rotational energy. For perpendicular moments ($c=0$), $J_{ij}^{\mathrm{proj}}=-h_{22}$; the other three components are not described by isotropic Heisenberg exchange.

The difference $\Delta H_{ij}=H_{ij}+J_{ij}^{\mathrm{proj}}P_{ij}$ measures the pair curvature not described by isotropic Heisenberg exchange. It does not identify the additional interactions. Even if $\Delta H_{ij}=0$, the agreement applies only to the off-site Hessian at the chosen magnetic configuration, not necessarily to torques, on-site curvatures, or finite-angle energies. Exchange coefficients obtained from different magnetic states or finite-angle fits may therefore differ. The exchange coefficients extracted from $H^B$ and $H^M$ can differ because the two curvatures describe different energy changes.

In collinear calculations, the sign of individual Hessian components depends on whether local rotation axes or common transverse displacement axes are used; the geometrical factor accounts for this convention.

A bilinear--biquadratic model illustrates how different interactions can contribute to the same local Hessian. Following the example discussed by dos Santos Dias et al.~\cite{ref25}, consider two unit-length spin directions with energy

\begin{equation}
E_{ij} = - J_{ij}\,\mathbf{e}_{i} \cdot \mathbf{e}_{j} + Q_{ij}\left( \mathbf{e}_{i} \cdot \mathbf{e}_{j} \right)^{2}.
\label{eq:biquad_pair_energy}
\end{equation}

The model contains no SOC or DMI. It shows how biquadratic exchange affects the Heisenberg projection in Eq.~\eqref{eq:hessian_to_j_projection}; related noncollinear dimer models are discussed in Ref.~\cite{ref27}.

We choose a reference state in which the two spins make an angle \(\phi\). This angle need not minimize the energy of the isolated pair: neighboring spins or external constraints may hold it fixed.

For infinitesimal transverse variations, \(\mathbf{e}_{i} \cdot \delta\mathbf{e}_{i} = \mathbf{e}_{j} \cdot \delta\mathbf{e}_{j} = 0\), the mixed second variation (one variation at each site) is

\begin{equation}
\begin{aligned}
\delta_{i}\delta_{j}E_{ij} = & \left( - J_{ij} + 2Q_{ij}\cos\phi \right)\,\delta\mathbf{e}_{i} \cdot \delta\mathbf{e}_{j} \\
 & + 2Q_{ij}\left( \delta\mathbf{e}_{i} \cdot \mathbf{e}_{j} \right)\left( \mathbf{e}_{i} \cdot \delta\mathbf{e}_{j} \right).
\end{aligned}
\label{eq:biquad_geometry_matrix}
\end{equation}

The standard vector identity gives, for these transverse variations,

\begin{equation}
\left( \mathbf{e}_{i} \times \mathbf{e}_{j} \right) \cdot \left( \delta\mathbf{e}_{i} \times \delta\mathbf{e}_{j} \right) = - \left( \mathbf{e}_{i} \cdot \delta\mathbf{e}_{j} \right)\left( \mathbf{e}_{j} \cdot \delta\mathbf{e}_{i} \right).
\label{eq:biquad_vector_identity}
\end{equation}

Consequently,

\begin{equation}
\begin{aligned}
\delta_{i}\delta_{j}E_{ij} = & \left( - J_{ij} + 2Q_{ij}\cos\phi \right)\,\delta\mathbf{e}_{i} \cdot \delta\mathbf{e}_{j} \\
 & - 2Q_{ij}\left( \mathbf{e}_{i} \times \mathbf{e}_{j} \right) \cdot \left( \delta\mathbf{e}_{i} \times \delta\mathbf{e}_{j} \right).
\end{aligned}
\label{eq:biquad_rotational_matrix}
\end{equation}

We can also examine the full pair curvature without assuming which magnetic interactions produce it. Using the transverse axes introduced above, with $c=\cos\phi$ and
$P_{ij}=\mathrm{diag}(c,1)$, the pair curvature is
\begin{equation}
H_{ij}=\begin{pmatrix}
-J_{ij}c+2Q_{ij}(2c^2-1)&0\\
0&-J_{ij}+2Q_{ij}c
\end{pmatrix}.
\label{eq:biquad_pair_hessian}
\end{equation}
The projected Heisenberg exchange coefficient and the part not described by it are
\begin{equation}
J_{ij}^{\mathrm{proj}}=J_{ij}-\frac{4Q_{ij}c^3}{1+c^2},
\label{eq:biquad_j_projection}
\end{equation}
\begin{equation}
\Delta H_{ij}=-\frac{2Q_{ij}(1-c^2)}{1+c^2}
\begin{pmatrix}1&0\\0&-c\end{pmatrix}.
\label{eq:biquad_j_residual}
\end{equation}
A biquadratic interaction therefore produces a pair curvature that generally cannot be represented by a single Heisenberg exchange coefficient. For parallel or antiparallel moments, however, its contribution to the transverse curvature has the Heisenberg form, with effective coefficients $J_{ij}-2Q_{ij}$ and $J_{ij}+2Q_{ij}$, respectively. At $c=0$, the exchange expression gives $J_{ij}$ but does not reproduce the full pair curvature when $Q_{ij}\ne0$. Therefore, agreement with the Heisenberg form for small rotations does not exclude higher-order interactions or guarantee that the same exchange parameters describe other magnetic configurations.

The second term looks like the mixed variation of a DMI energy \(\mathbf{D}_{ij} \cdot \left( \mathbf{e}_{i} \times \mathbf{e}_{j} \right)\). If interpreted in isolation, it would suggest the \textbf{apparent}, reference-dependent coefficient

\begin{equation}
\mathbf{D}_{ij}^{app} = - 2Q_{ij}\left( \mathbf{e}_{i} \times \mathbf{e}_{j} \right),
\label{eq:apparent_dm}
\end{equation}

with magnitude \(2|Q_{ij}\sin\phi|\). This apparent DMI depends on the reference angle. It is not a separate chiral interaction: the same biquadratic term also changes the coefficient of \(\delta\mathbf e_i\cdot\delta\mathbf e_j\). Both contributions belong to one Hessian.

We can check that the model has no DMI by reversing the relative angle. For coplanar spins,

\begin{equation}
E_{ij}(\phi) = - J_{ij}\cos\phi + Q_{ij}\cos^{2}\phi = E_{ij}( - \phi).
\label{eq:biquad_even_energy}
\end{equation}

Opposite chiralities have the same energy, although the apparent DMI coefficient changes sign. The antisymmetric-looking Hessian term therefore does not reveal a physical DMI. This issue is separate from the difference between field and moment rotations discussed in Sec.~\ref{sec:pt_moment}.

Infinitesimal rotations probe only directions perpendicular to each reference spin. They cannot determine longitudinal components of an exchange tensor. Thus, different exchange tensors can give the same transverse energy derivatives. These derivatives are uniquely determined, but their interpretation in terms of bilinear, symmetric anisotropic, and DMI-like interactions is not generally unique.

A three-spin example shows how the apparent interaction between two spins can depend on a third. We consider the rotationally invariant energy

\begin{equation}
E =  - J(\mathbf{e}_{1} \cdot \mathbf{e}_{2}\  + \ \mathbf{e}_{2} \cdot \mathbf{e}_{3})\  - \ K_{\mathrm{3s}}(\mathbf{e}_{1} \cdot \mathbf{e}_{2})(\mathbf{e}_{2} \cdot \mathbf{e}_{3}).
\label{eq:three_spin_energy}
\end{equation}

For coplanar spins, define $\alpha=\theta_1-\theta_2$ and $\beta=\theta_2-\theta_3$. The mixed curvature between the end spins, despite the absence of a bilinear $(\mathbf e_1\cdot\mathbf e_3)$ term, is

\begin{equation}
\partial^{2}E/(\partial\theta_{1}\partial\theta_{3}) = K_{\mathrm{3s}}\sin\alpha\sin\beta.
\label{eq:three_spin_mixed}
\end{equation}

For a fixed orientation of spin 3, the coefficient of
$(\mathbf e_1\cdot\mathbf e_2)$ becomes $J+K_{\mathrm{3s}}\cos\beta$,
demonstrating its dependence on the orientation of the third spin.
The three-spin interaction also contributes directly to the
transverse coupling between spins 1 and 3. For coplanar reference
directions $\mathbf e_i=(\cos\theta_i,\sin\theta_i,0)$, the local
transverse vectors are $\mathbf t_i^1=(-\sin\theta_i,\cos\theta_i,0)$ and
$\mathbf t_i^2=(0,0,1)$. The mixed second variation is
\begin{equation}
\delta_1\delta_3 E=-K_{\mathrm{3s}}(\delta\mathbf e_1\cdot\mathbf e_2)
(\mathbf e_2\cdot\delta\mathbf e_3),
\label{eq:three_spin_mixed_variation}
\end{equation}
where the bilinear terms do not contribute to the coupling between spins 1 and 3.
Writing $s=\sin\alpha\sin\beta$ and
$c=\mathbf e_1\cdot\mathbf e_3=\cos(\alpha+\beta)$ gives
\begin{equation}
H_{13}=\begin{pmatrix}K_{\mathrm{3s}}s&0\\0&0\end{pmatrix},\qquad
P_{13}=\begin{pmatrix}c&0\\0&1\end{pmatrix}.
\label{eq:three_spin_full_block}
\end{equation}
Consequently, Eq.~\eqref{eq:hessian_to_j_projection} yields
\begin{equation}
J_{13}^{\mathrm{proj}}=-\frac{K_{\mathrm{3s}}cs}{1+c^2},\qquad
\Delta H_{13}=\frac{K_{\mathrm{3s}}s}{1+c^2}
\begin{pmatrix}1&0\\0&-c\end{pmatrix}.
\label{eq:three_spin_projection}
\end{equation}
For $K_{\mathrm{3s}}s\ne0$, the relative Frobenius difference is
\begin{equation}
r_{13}=\frac{1}{\sqrt{1+c^2}},\qquad
\frac{1}{\sqrt{2}}\le r_{13}\le1.
\label{eq:three_spin_residual}
\end{equation}
For nonzero $K_{\mathrm{3s}}s$, at least half the squared magnitude of the pair curvature lies outside the isotropic Heisenberg form. At $c=0$ the projected coefficient
vanishes even though $H_{13}$ can remain finite; at $s=0$ the pair curvature between spins 1 and 3 vanishes, so its relative difference is undefined.
Thus, the exchange coefficient inferred from the local curvature need not represent a bilinear interaction between spins 1 and 3. Because this comparison concerns only the local curvature, it does not establish the stability of the reference state.

\subsection{Comparison with earlier noncollinear formulations}

To compare with other noncollinear calculations, we first express the response in spin components:

\begin{equation}
\mathcal{A}_{ij}^{\alpha\beta} = - \frac{1}{\pi}\operatorname{Im}\int_{}^{E_{F}}{d\varepsilon\, \operatorname{Tr}\left\lbrack b_{i}\sigma_{\alpha}G_{ij}b_{j}\sigma_{\beta}G_{ji} \right\rbrack}
\label{eq:orig041}
\end{equation}
\begin{equation}
H_{ij}^{ab} = \mathbf{t}_{i\alpha}^{a}\mathcal{A}_{ij}^{\alpha\beta}\mathbf{t}_{j\beta}^{b}
\label{eq:orig042}
\end{equation}
Reference \cite{ref19} obtained the energy of prescribed rotations in a general magnetic configuration using scattering matrices. Its formulation introduces electronic response coefficients and a noncollinear interaction tensor, together with local contributions and rotational constraints. In the notation of Ref.~\cite{ref19}, the proposed spin-interaction tensor is \(J_{A,ij}^{\alpha\beta}=A_{ij}^{00}\delta_{\alpha\beta}-A_{ij}^{\alpha\beta}\), subject to that work's scattering definitions and pair-counting convention. Comparison with Eq.~\eqref{eq:orig042} requires the second derivative with respect to spin rotations, including the local terms and the dependence on the reference directions.

The comparison with Ref.~\cite{ref19} therefore depends on the rotational energy change, not just the names of the tensor components. Ref.~\cite{ref26} instead uses coefficients \(A_{ij}^{\mu\nu}\) constructed from scalar and spin-vector components of the scattering-path operator. They differ from \(\mathcal{A}_{ij}^{\alpha\beta}\) in Eq.~\eqref{eq:orig041}. Their connected two-site energy variation contains the symmetric combination

\begin{equation}
K_{S,ij}^{(\alpha\beta)} = \left( A_{ij}^{00} - \sum_{\mu = x,y,z}^{}A_{ij}^{\mu\mu} \right)\delta_{\alpha\beta} + 2A_{ij}^{(\alpha\beta)}
\label{eq:orig043}
\end{equation}

Here the coefficient \(K_S\) denotes the term inside the published two-site energy variation, before applying the overall prefactor and pair-counting convention. In contrast, the spin-energy construction involves \(A_{ij}^{00}\ \delta_{\alpha\beta}\  - \ A_{ij}^{\alpha\beta}\), where \(A_{ij}^{00}\) and \(A_{ij}^{\alpha\beta}\) denote the spin-scalar and spin-dependent components, respectively. These expressions describe different quantities. A proposed bilinear spin energy must reproduce both the two-site and one-site energy changes. Its antisymmetric terms and rotational sum rule must also agree. Matching the two-site expressions does not by itself show that the tensors are equal. The Green-function product in Eq.~\eqref{eq:orig039} can be compared with a scattering-path expression only after matching the vertices, propagators, and pair-counting conventions.

\subsection{Collinear limit}

For moments along $z$, Eq.~\eqref{eq:orig039} reduces to the exchange formula of Refs.~\cite{ref13,ref16,ref17,ref18}. For the half-sum pair convention specified above, the transverse response gives

\begin{equation}
J_{ij}^{B} = \frac{1}{\pi}\operatorname{Im}\int_{}^{E_{F}}{d\varepsilon\,\operatorname{Tr}_{L}\left\lbrack b_{i}G_{ij}^{\uparrow}b_{j}G_{ji}^{\downarrow} + b_{i}G_{ij}^{\downarrow}b_{j}G_{ji}^{\uparrow} \right\rbrack}
\label{eq:orig044}
\end{equation}
In the scattering notation of Ref. \cite{ref26}, the corresponding collinear combination is \(J_{ij}=A_{ij}^{00}-A_{ij}^{zz}\), with their normalization.

The original PK band expression and the LKG multiple-scattering formula are shown to give the same exchange kernel in Appendix~\ref{app:pk_lkg}, with the corresponding scattering-state normalization and pair-counting conventions.

\subsection{Collinear exchange with a nonlocal magnetic potential}
\label{sec:nonlocalstationary}

Equation~\eqref{eq:orig044} assumes a site-local magnetic potential. For a nonlocal operator, we must specify which part of the Hamiltonian rotates and how it changes under site rotations.

We write the spin-diagonal reference Hamiltonian as $H_0^s=h_0+\Sigma^s$ ($s=\uparrow,\downarrow$), where $\Sigma^s$ is a matrix in the orbital and site indices. We include all spin-independent terms, local or nonlocal, in $h_0$ and take $\Sigma^s$ as the magnetic operator to be rotated. We let $\mathcal P_i$ project onto the orbitals whose spin direction is rotated at site $i$. A prescribed local rotation of the \emph{entire} nonlocal magnetic operator is defined by $\Sigma(\boldsymbol\theta)=U(\boldsymbol\theta)\Sigma_0U^\dagger(\boldsymbol\theta)$, where $U=\exp[-i\sum_i\theta_i\mathcal P_i\sigma_y/2]$. Its first-order transverse vertex has the spin-flip blocks
\begin{align}
 V_i^{\uparrow\downarrow}&=\tfrac12(\Sigma^\uparrow \mathcal P_i-\mathcal P_i\Sigma^\downarrow),\nonumber\\
 V_i^{\downarrow\uparrow}&=\tfrac12(\mathcal P_i\Sigma^\uparrow-\Sigma^\downarrow \mathcal P_i).
 \label{eq:nlvertex}
\end{align}
The unitary also rotates spin-dependent bond matrix elements. Moment directions alone do not determine how these elements rotate. Consequently, $J_{ij}^{B}$ depends on this choice. We use mutually orthogonal, complete site projectors, $\mathcal P_i\mathcal P_j=\delta_{ij}\mathcal P_i$ and $\sum_i\mathcal P_i=I$.

For a local splitting $\Sigma^\uparrow=-\sum_i b_i\mathcal P_i$ and $\Sigma^\downarrow=+\sum_i b_i\mathcal P_i$, both blocks reduce to $-b_i\mathcal P_i$, as in Eq.~\eqref{eq:orig044}. For a nonlocal operator, $V_i$ also connects site $i$ to other sites. These terms are missed if one merely replaces $b_i$ by a nonlocal splitting in Eq.~\eqref{eq:orig044}.

For these rotations, the pair curvature includes the Green-function contribution from the nonlocal vertices and the mixed second-order contact term. With $J_{ij}^{B}=-H_{ij}^{B}$, the result is
\begin{align}
 J_{ij}^{B}={}&\frac{1}{\pi}\operatorname{Im}\int^{E_F}\!d\varepsilon\,\operatorname{Tr}\bigl[
 V_i^{\uparrow\downarrow}G^\downarrow
 V_j^{\downarrow\uparrow}G^\uparrow\nonumber\\
 &+V_i^{\downarrow\uparrow}G^\uparrow
 V_j^{\uparrow\downarrow}G^\downarrow\bigr]
 -\operatorname{Tr}(\gamma_0W_{ij}),\quad i\ne j.
 \label{eq:nlGF}
\end{align}
Here the trace is over all orbital and site indices, $G^s=(\varepsilon-H_0^s+i0^+)^{-1}$, and $W_{ij}=\partial_{\theta_i}\partial_{\theta_j}\Sigma(\boldsymbol\theta)|_0$ is the mixed second-order vertex. In the choice of magnetic-operator rotation above,
\begin{equation}
 W_{ij}=-\tfrac14[\mathcal P_i\sigma_y,[\mathcal P_j\sigma_y,\Sigma_0]],
 \qquad i\ne j.
 \label{eq:nlcontact}
\end{equation}
The contact term vanishes for a site-diagonal magnetic potential at distinct sites; it need not vanish for a nonlocal potential. Equation~\eqref{eq:nlGF} therefore reduces to Eq.~\eqref{eq:orig044} in the local limit (with the same pair-counting convention). It is the prescribed-rotation curvature of the corresponding one-electron spectral energy; its identification with a total-energy curvature requires an appropriate functional and the same rotation path.

The contact contribution, including its mixed nonlocal term, is required for the cancellation of the uniform-rotation second derivative of the magnetic energy. For a uniform rotation generated by $\mathcal S=\sigma_y/2$, with the spin-independent Hamiltonian commuting with $\mathcal S$, the reference Hamiltonian transforms as $H(\Theta)=e^{-i\Theta \mathcal S}H_{\rm ref}e^{i\Theta \mathcal S}$. The corresponding perturbations are $V=-i[\mathcal S,H_{\rm ref}]$ and $W=-[\mathcal S,[\mathcal S,H_{\rm ref}]]$. In the eigenbasis of $H_{\rm ref}$, the contact and bubble contributions satisfy
\begin{equation}
 \langle W\rangle
 =2\sum_{h\in\mathrm{occ},\,p\in\mathrm{emp}}
 (\varepsilon_p-\varepsilon_h)|\mathcal S_{ph}|^2
 =-H_{\rm bubble}^{\rm unif},
 \label{eq:stage2OperatorWard}
\end{equation}
The total uniform-rotation curvature therefore vanishes, as required by spin-rotational symmetry. This cancellation involves all site contributions, not individual pair interactions $J_{ij}^{B}$, and requires the complete first- and second-order perturbations generated by the same rotation. Omitting the mixed nonlocal contact term can violate this condition.

For moment rotations, a different curvature follows from the specified stationary functional $E[\gamma]=\operatorname{Tr}(h_0\gamma)+\Phi_{\mathrm{stat}}[\gamma]$. This result assumes a differentiable energy functional and a stationary reference. It does not apply automatically to arbitrary nonlocal operators. The interaction functional $\Phi$ retains all interaction contributions, including nonlocal terms, even when these terms do not appear explicitly in the prescribed one-electron rotation vertices. The reference potential and the induced potential are, respectively,
\begin{equation}
 \Sigma_0=\left.\frac{\delta\Phi_{\mathrm{stat}}}{\delta\gamma}\right|_{\gamma_0},
 \qquad \delta\Sigma=\mathcal I(\delta\gamma),
 \label{eq:nlkernel}
\end{equation}
where $\mathcal I$ is the second functional derivative of $\Phi_{\mathrm{stat}}$. For an infinitesimal rotation of the \emph{constrained moment} at site $i$, we denote the self-consistent first-order density-matrix change by $\delta_i\gamma$. In the occupied--empty eigenstate basis of $H_0$, we write $\delta_i\gamma=\sum_{ph}(z_{ph}^{(i)}|p\rangle\langle h|+\mathrm{h.c.})$. This expansion in occupied and empty states assumes a zero-temperature reference density matrix with $\gamma^2=\gamma$. The stationary mixed second derivative follows directly from the second-order energy:
\begin{align}
 J_{ij}^{M}=-{}&2\operatorname{Re}\sum_{ph}
 (\varepsilon_p-\varepsilon_h)
 z_{ph}^{(i)*}z_{ph}^{(j)}\nonumber\\
 &-\operatorname{Tr}\bigl[\delta_i\gamma\,
 \mathcal I(\delta_j\gamma)\bigr],\qquad i\ne j.
 \label{eq:nlJstationary}
\end{align}
We determine the amplitudes $z_{ph}^{(i)}$ from stationarity of the same quadratic energy subject to the prescribed transverse moments, $\operatorname{Tr}(\mathcal P_k\sigma_x\delta_i\gamma)=M_k\delta_{ik}$, with the sign of $M_k$ understood in the chosen local transverse coordinates. This specifies the constraint on the transverse $x$ component; the $y$ channel follows by spin-rotation symmetry. All other electronic amplitudes are allowed to relax. Equation~\eqref{eq:nlJstationary} gives the constrained-moment curvature for this stationary functional: it retains the particle--hole contribution and the explicit change of the interaction energy. In general, the second term cannot be expressed using only a site exchange splitting or a product of independent-particle Green functions.

The amplitudes in Eq.~\eqref{eq:nlJstationary} must minimize the quadratic energy while satisfying the moment constraints. All other allowed particle--hole excitations must be included in the electronic response. Suppressing those excitations or omitting the induced interaction $\mathcal I$ gives a different second derivative, not the stated $J_{ij}^{M}$.

The stationary functional has its own, distinct Goldstone cancellation. If $h_0$ is spin independent and $\Phi[R\gamma R^\dagger]=\Phi_{\mathrm{stat}}[\gamma]$ under a global spin rotation, the uniform tangent $\delta\gamma_{\rm unif}=-i[\mathcal S,\gamma_0]$ is an exact symmetry direction. Evaluating the stationary second variation along this direction gives
\begin{equation}
 2\sum_{ph}(\varepsilon_p-\varepsilon_h)
 |z_{ph}^{\rm unif}|^2
 +\operatorname{Tr}\!\left[\delta\gamma_{\rm unif}\,
 \mathcal I(\delta\gamma_{\rm unif})\right]=0.
 \label{eq:stage2FunctionalWard}
\end{equation}
Here $z_{ph}^{\rm unif}$ denotes the occupied--empty amplitude of the global rotation. The induced functional-kernel contribution cancels the positive particle--hole term; the cancellation is not obtained in general by keeping only the equilibrium nonlocal potential. This is the stationary-total-energy identity, not the prescribed-potential contact-term identity~\eqref{eq:stage2OperatorWard}.

These expressions describe different rotations: Eq.~\eqref{eq:nlGF} gives the prescribed-potential response for a nonlocal operator, whereas Eq.~\eqref{eq:nlJstationary} gives the constrained-moment exchange from the stationary total energy. Their difference can involve electronic relaxation, changes in the interaction, and the different constraints and energy functionals discussed in Sec.~\ref{sec:pt_moment}. Without a derivation from the same functional, it cannot be assigned to a single relaxation correction. The two-electron example in Appendix~\ref{app:twosite} evaluates the stationary result explicitly and finds
\begin{equation}
 J_{AB}^{M}=-\frac{2t^2}{2U-K_{\mathrm{nl}}},\qquad 2U-K_{\mathrm{nl}}>4t.
 \label{eq:nltwositeJ}
\end{equation}
Here $K_{\mathrm{nl}}$ multiplies an orbital-nonlocal density-matrix interaction, not a site-only Stoner coupling. This model result illustrates why the full nonlocal functional must be retained; it is not a general replacement $U\mapsto U-K_{\mathrm{nl}}/2$.

A generalized Stoner functional illustrates the difference between a nonlocal one-electron magnetic potential and a nonlocal response kernel:
\begin{equation}
 F[\gamma]=F_0[\gamma]-\frac14\sum_{ij}I_{ij}\mathbf M_i\cdot\mathbf M_j,
 \label{eq:nlStonerFunctional}
\end{equation}
which generates a site-local magnetic field $\mathbf B_i=\frac12\sum_j I_{ij}\mathbf M_j$, even when $I_{ij}$ is nonlocal. For a translationally invariant collinear ferromagnet, $B=I(0)M/2$. In terms of the independent-electron static transverse susceptibility $\chi_0(q)$, defined by $\delta M_\perp(q)=\chi_0(q)\delta B_\perp(q)$ and satisfying $\chi_0(0)=M/B$, the difference between the constrained-moment and prescribed-field rotational energy derivatives takes the form
\begin{equation}
\begin{aligned}
 H^{M}(q)-H^{B}(q)={}&\frac{[M-B\chi_0(q)]^2}{\chi_0(q)}\\
 &+\frac{M^2}{2}[I(0)-I(q)].
\end{aligned}
\label{eq:nlStonerDifference}
\end{equation}
Here $H^{B}=B^2[\chi_0(0)-\chi_0(q)]$ is the curvature obtained by rotating the frozen equilibrium one-electron field, whereas $H^{M}$ is the stationary constrained-moment curvature of Eq.~\eqref{eq:nlStonerFunctional}; the expression assumes a regular transverse Legendre transform. Rotating the input magnetization in the full nonlocal interaction functional defines a different energy variation and must not be identified with this $H^{B}$. In the Pauli-spin response convention used here, the Stoner functional generates $\delta B(q)=[I(q)/2]\delta M(q)$, so the interaction kernel entering the inverse transverse susceptibility is $I_{\mathrm{resp}}(q)=I(q)/2$. For an analytic, inversion-symmetric response, the first term begins at order $q^4$, whereas the second can contribute already at order $q^2$. Writing $I(0)-I(\mathbf q)=K_{\alpha\beta}q_\alpha q_\beta+O(q^4)$ gives the explicit leading-curvature difference
\begin{equation}
A^M_{\alpha\beta}-A^B_{\alpha\beta}=\frac{M^2}{2}K_{\alpha\beta}.
\label{eq:nl_stiffness_difference}
\end{equation}

\section{MAGNETIC PAIR INTERACTIONS IN SPIN SPIRALS}

The pair interactions derived in Sec.~\ref{sec:pt_interactions} depend on the magnetic configuration. Planar and conical spin spirals allow us to examine whether a single isotropic exchange coefficient describes rotations about different noncollinear states. Here $\vartheta$ denotes the cone angle, \(\mathbf Q\) the spiral wave vector, and \(\mathbf R_i\) a site position; Eq.~\eqref{eq:orig046} specifies the site phase.

\begin{equation}
\mathbf{e}_{i} = \left( \sin\vartheta\cos\phi_{i},\,\sin\vartheta\sin\phi_{i},\,\cos\vartheta \right)
\label{eq:orig045}
\end{equation}
with

\begin{equation}
\phi_{i} = \mathbf{Q} \cdot \mathbf{R}_{i}
\label{eq:orig046}
\end{equation}
A convenient orthonormal tangent basis is

\begin{equation}
\mathbf{t}_{i}^{1} = \left( \cos\vartheta\cos\phi_{i},\,\cos\vartheta\sin\phi_{i},\, - \sin\vartheta \right)
\label{eq:orig047}
\end{equation}
and

\begin{equation}
\mathbf{t}_{i}^{2} = \left( - \sin\phi_{i},\,\cos\phi_{i},\, 0 \right)
\label{eq:orig048}
\end{equation}
The labels 1 and 2 denote rotations along the two transverse directions. A change of spiral phase instead moves the moment by

\begin{equation}
\frac{\partial\mathbf{e}_{i}}{\partial\phi_{i}} = \sin\vartheta\,\mathbf{t}_{i}^{2}
\label{eq:orig049}
\end{equation}
Below, \(\theta_{i}^{2}\) measures the transverse rotation, not the change in spiral phase. The two differ by a factor of $\sin\vartheta$ in a conical spiral.

For the spiral tangent vectors of Eqs.~\eqref{eq:orig047}, \eqref{eq:orig048}, the first-order vertices are given by Eq.~\eqref{eq:orig008}, and the off-site electronic pair Hessian by Eq.~\eqref{eq:orig016}.

\subsection{Conical spiral: derivation of scalar pair interactions}

The pair curvature of the isotropic Heisenberg model in Eq.~\eqref{eq:orig018} gives the geometrical factor for a conical spiral.

Writing \(\phi_{ij}=\phi_i-\phi_j=\mathbf Q\cdot(\mathbf R_i-\mathbf R_j)\), direct evaluation for the conical spiral gives

\begin{equation}
P_{ij}=\begin{pmatrix}
\sin^{2}\vartheta+\cos^{2}\vartheta\cos\phi_{ij} & \cos\vartheta\sin\phi_{ij}\\
-\cos\vartheta\sin\phi_{ij} & \cos\phi_{ij}
\end{pmatrix}
\label{eq:orig051}
\end{equation}

A scalar bilinear interaction can reproduce the electronic rotational response only if

\begin{equation}
H_{ij}^{ab} = - J_{ij}P_{ij}^{ab}
\label{eq:orig052}
\end{equation}
Equation~\eqref{eq:orig052} is the condition for describing the calculated pair response by isotropic bilinear exchange.

The two diagonal transverse directions give the following exchange coefficients when their geometrical factors are nonzero:

\begin{equation}
J_{ij}^{(1)} = - \frac{H_{ij}^{11}}{\sin^{2}\vartheta + \cos^{2}\vartheta\cos\phi_{ij}}
\label{eq:orig053}
\end{equation}
and

\begin{equation}
J_{ij}^{(2)} = - \frac{H_{ij}^{22}}{\cos\phi_{ij}}
\label{eq:orig054}
\end{equation}
Agreement between these two values is not sufficient: the mixed transverse responses must also have the Heisenberg form. In a conical spiral, the mixed channels also obey

\begin{equation}
H_{ij}^{12} = - J_{ij}\cos\vartheta\sin\phi_{ij}
\label{eq:orig055}
\end{equation}
and

\begin{equation}
H_{ij}^{21} = J_{ij}\cos\vartheta\sin\phi_{ij}
\label{eq:orig056}
\end{equation}
The opposite signs follow from the symmetry of the complete Hessian,

\begin{equation}
H_{ij}^{ab} = H_{ji}^{ba}
\label{eq:orig057}
\end{equation}
Equations~\eqref{eq:orig053}--\eqref{eq:orig057} test diagonal and mixed channels. Changing the cone angle or spiral wave vector shows whether the same exchange coefficient applies to other magnetic configurations.

\subsection{Planar spin spiral}

In a planar spiral, the tangent geometry simplifies:

\begin{equation}
\vartheta = \frac{\pi}{2}
\label{eq:orig058}
\end{equation}
and the tangent vectors become

\begin{equation}
\mathbf{t}_{i}^{1} = (0,0, - 1),\quad\quad\mathbf{t}_{i}^{2} = \left( - \sin\phi_{i},\cos\phi_{i},0 \right)
\label{eq:orig059}
\end{equation}
The geometrical matrix is then diagonal,

\begin{equation}
P_{ij}=\begin{pmatrix}1&0\\0&\cos\phi_{ij}\end{pmatrix}
\label{eq:orig060}
\end{equation}

Here channel 1 denotes the out-of-plane tilt, whereas channel 1 in the local frame of Sec.~\ref{sec:pt_noncollinear} denotes the in-plane tilt; the channel labels are local conventions, not different physical responses. The out-of-plane rotation gives a simpler expression. A scalar exchange would require

\begin{equation}
H_{ij}^{11} = - J_{ij}
\label{eq:orig061}
\end{equation}
and hence

\begin{equation}
J_{ij}^{B} = - H_{ij}^{11}
\label{eq:orig062}
\end{equation}
This expression contains no angle-dependent denominator and is valid for every relative spin angle.

The out-of-plane electronic vertex follows from Eq.~\eqref{eq:orig008} using the tangent convention of Eq.~\eqref{eq:orig059}:

\begin{equation}
V_{i}^{1} = b_{i}\sigma_{z}
\label{eq:orig063}
\end{equation}
Substitution into Eq.~\eqref{eq:orig016} gives

\begin{equation}
H_{ij}^{11} = - \frac{b_{i}b_{j}}{2\pi}\operatorname{Im}\int_{}^{E_{F}}{d\varepsilon \operatorname{Tr}\left\lbrack \sigma_{z}G_{ij}\sigma_{z}G_{ji} + \sigma_{z}G_{ji}\sigma_{z}G_{ij} \right\rbrack}
\label{eq:orig064}
\end{equation}
and therefore

\begin{equation}
J_{ij}^{B} = \frac{b_{i}b_{j}}{2\pi}\operatorname{Im}\int_{}^{E_{F}}{d\varepsilon \operatorname{Tr}\left\lbrack \sigma_{z}G_{ij}\sigma_{z}G_{ji} + \sigma_{z}G_{ji}\sigma_{z}G_{ij} \right\rbrack}
\label{eq:orig065}
\end{equation}
The in-plane and mixed channels provide independent checks. For the in-plane channel, Eq.~\eqref{eq:orig052} requires

\begin{equation}
H_{ij}^{22} = - J_{ij}\cos\phi_{ij}
\label{eq:orig066}
\end{equation}
or, equivalently,

\begin{equation}
H_{ij}^{22} = H_{ij}^{11}\cos\phi_{ij}
\label{eq:orig067}
\end{equation}
Unlike a determination based on division by $\cos\phi_{ij}$, Eq.~\eqref{eq:orig067} remains regular when $\cos\phi_{ij}=0$.

For a scalar isotropic pair interaction, the mixed components must satisfy

\begin{equation}
H_{ij}^{12} = H_{ij}^{21} = 0
\label{eq:orig068}
\end{equation}
Equations~\eqref{eq:orig062} and \eqref{eq:orig065} define the scalar coefficient through the out-of-plane channel; Eqs.~\eqref{eq:orig067} and \eqref{eq:orig068} are then the independent tests, and both must hold. In particular, planar geometry alone does not make the mixed electronic responses vanish.

For the isotropic bilinear--biquadratic pair of Sec.~\ref{sec:pt_noncollinear}, the planar-spiral transverse curvatures provide an explicit test:

\begin{equation}
H_{ij}^{\parallel} = \frac{\partial^{2}E_{ij}}{\partial\alpha_{i}\partial\alpha_{j}} = - J_{ij}\cos\phi + 2Q_{ij}\cos(2\phi).
\label{eq:biquad_h_parallel}
\end{equation}

For small out-of-plane tilts \(\beta_{i}\) and \(\beta_{j}\) about the same planar reference, it is

\begin{equation}
H_{ij}^{\bot} = \frac{\partial^{2}E_{ij}}{\partial\beta_{i}\partial\beta_{j}} = - J_{ij} + 2Q_{ij}\cos\phi.
\label{eq:biquad_h_perp}
\end{equation}

A purely bilinear isotropic Heisenberg pair with scalar coefficient \(J_{ij}^{eff}\) would instead give \(H_{ij}^{\parallel ,H} = - J_{ij}^{eff}\cos\phi\) and \(H_{ij}^{\bot,H} = - J_{ij}^{eff}\). Thus the two independent determinations give

\begin{equation}
J_{ij}^{eff, \parallel} = J_{ij} - 2Q_{ij}\frac{\cos(2\phi)}{\cos\phi},\qquad \cos\phi\ne 0,
\label{eq:biquad_jeff_parallel}
\end{equation}

\begin{equation}
J_{ij}^{eff,\bot} = J_{ij} - 2Q_{ij}\cos\phi,
\label{eq:biquad_jeff_perp}
\end{equation}

and consequently

\begin{equation}
J_{ij}^{eff, \parallel} - J_{ij}^{eff,\bot} = \frac{2Q_{ij}\sin^{2}\phi}{\cos\phi}.
\label{eq:biquad_jeff_difference}
\end{equation}

Equation~\eqref{eq:biquad_jeff_difference} shows that for a general noncollinear angle and nonzero \(Q_{ij}\), the two exchange coefficients differ. At \(\phi = \pi/2\), the in-plane Heisenberg curvature vanishes for \emph{any finite} scalar \(J_{ij}^{eff}\), whereas the biquadratic model gives \(H_{ij}^{\parallel} = - 2Q_{ij}\). Thus, a single scalar bilinear interaction cannot reproduce both channels at this reference configuration. The apparent singularity of Eq.~\eqref{eq:biquad_jeff_parallel} at \(\cos\phi = 0\) arises because the bilinear geometrical factor vanishes. The original energy curvature remains finite.

No longitudinal tensor components enter this comparison.

For this coplanar bilinear--biquadratic model, the mixed in-plane/out-of-plane curvatures vanish by symmetry. This property does not hold for a general electronic pair response. The separate conditions are given in Eqs.~\eqref{eq:orig067} and \eqref{eq:orig068}, with the coefficient defined in Eq.~\eqref{eq:orig065}.

\subsection{Collinear limits}

The collinear ferromagnetic (FM) and two-sublattice antiferromagnetic (AFM) states are limiting cases of the planar spiral.

For

\begin{equation}
\mathbf{Q} = 0
\label{eq:orig069}
\end{equation}
all moments are parallel,

\begin{equation}
\mathbf{e}_{i} = \mathbf{e}_{j}
\label{eq:orig070}
\end{equation}
For the nonrelativistic isotropic problem, the two transverse channels are equivalent, and Eq.~\eqref{eq:orig052} becomes

\begin{equation}
H_{ij}^{11} = H_{ij}^{22} = - J_{ij},\quad\quad H_{ij}^{12} = H_{ij}^{21} = 0
\label{eq:orig071}
\end{equation}
This gives the familiar FM pair response; its long-wavelength limit follows in Sec.~\ref{sec:pt_fm}.

A collinear two-sublattice AFM is obtained from a commensurate spiral vector \(\mathbf Q_{\mathrm{AF}}\) satisfying

\begin{equation}
\mathbf Q_{\mathrm{AF}}\cdot(\mathbf R_i-\mathbf R_j)=\pi\pmod{2\pi}\quad\text{(opposite sublattices)}
\label{eq:orig072}
\end{equation}
For the planar representation,

\begin{equation}
\phi_{ij} = \pi
\label{eq:orig073}
\end{equation}
for an opposite-sublattice pair. Since

\begin{equation}
\cos\phi_{ij} = - 1,\quad\quad\sin\phi_{ij} = 0
\label{eq:orig074}
\end{equation}
the geometrical factors become

\begin{equation}
P_{ij}=\begin{pmatrix}1&0\\0&-1\end{pmatrix}
\label{eq:orig075}
\end{equation}

A scalar bilinear interaction therefore gives

\begin{equation}
H_{ij}^{11} = - J_{ij}
\label{eq:orig076}
\end{equation}
and

\begin{equation}
H_{ij}^{22} = + J_{ij}
\label{eq:orig077}
\end{equation}
The two AFM transverse channels have opposite signs because the local axes point in opposite directions. Both come from the same electronic pair response.

\section{FERROMAGNETIC LIMIT AND SPIN-WAVE STIFFNESS}
\label{sec:pt_fm}

We compare field and moment rotations in a ferromagnet and examine when their curvatures give the physical spin-wave stiffness.

The two transverse directions are equivalent. The two-vertex term gives the intersite energy curvature, while the rotational sum rule fixes the on-site term. We denote the off-site transverse Hessian elements by \(H_{0R}\).

\begin{equation}
\begin{aligned}
H_{0R} &=-\frac{1}{2\pi}\operatorname{Im}\int^{E_F}d\varepsilon\,\operatorname{Tr}\bigl[V_0G_{0R}V_RG_{R0}\\ &\qquad +V_RG_{R0}V_0G_{0R}\bigr],\qquad R\ne0
\end{aligned}
\label{eq:orig078}
\end{equation}
Expanding at small wave vector gives

\begin{equation}
H\left( \mathbf{q} \right) = \sum_{R\ne 0}^{}{H_{0R}\left( \mathbf{e}^{i\mathbf{q} \cdot \mathbf{R}} - 1 \right)} = A_{\alpha\beta}q_{\alpha}q_{\beta} + O\left( \mathbf{q}^{4} \right)
\label{eq:orig079}
\end{equation}
\begin{equation}
A_{\alpha\beta} = - \frac{1}{2}\sum_{R\ne 0}^{}{H_{0R}R_{\alpha}R_{\beta}}
\label{eq:orig080}
\end{equation}
Since $H_{0R}=H_{0,-R}$, odd powers of $q$ vanish. The coefficient $A$ in Eq.~\eqref{eq:orig080} is a coefficient of the long-wavelength rotational energy, distinct from the electronic response tensor $\mathcal{A}$ in Eq.~\eqref{eq:orig041}. We denote the equilibrium spin angular momentum per magnetic site, in units of $\hbar$, by $S$. Using the prescribed-field curvature and assuming an adiabatic precession coefficient $-S$ gives the stiffness estimate

\begin{equation}
\begin{aligned}
D^B_{\alpha\beta}&=\frac{A^B_{\alpha\beta}}{S}=\frac{1}{4\pi S}\operatorname{Im}\int^{E_F}d\varepsilon\sum_{R\ne0}R_\alpha R_\beta\\ &\qquad\times\operatorname{Tr}\bigl[V_0G_{0R}V_RG_{R0}+V_RG_{R0}V_0G_{0R}\bigr]
\end{aligned}
\label{eq:orig081}
\end{equation}
Equation~\eqref{eq:orig081} defines the prescribed-field stiffness coefficient. To identify it with the physical spin-wave stiffness, we must also examine the constrained-moment curvature and the precessional normalization.

The Pauli-spin convention of Eq.~\eqref{eq:orig001} is used here: the magnetic Hamiltonian is \(- b\bm{\sigma} \cdot \mathbf{e}\), with \(b = \frac{\Delta}{2}\) and \(m = \langle\sigma_{z}\rangle\) per site. The bare transverse response \(X_{0}(\mathbf{q})\) is defined by \(\delta\langle\sigma_{x}\rangle = X_{0}(\mathbf{q})\delta b_{x}\) for the perturbation \(- \delta b_{x}\sigma_{x}\), with the same site projection and Fourier convention as the rotational vertices. The local second-order rotational term contributes \(+ bm\) to the complete second derivatives with respect to rotation angles, whereas the two-vertex response contributes \(- b^{2}X_{0}(\mathbf{q})\). Their cancellation for a uniform rotation gives \(m = bX_{0}(0)\). Thus the complete field Hessian, not merely its off-site pair term, is

\begin{equation}
H^{B}(\mathbf{q}) = bm\  - \ b^{2}X_{0}(\mathbf{q}) = b^{2}\lbrack X_{0}(0)\  - \ X_{0}(\mathbf{q})\rbrack.
\label{eq:orig082}
\end{equation}
For the relaxed moment, Eqs.~\eqref{eq:orig026}--\eqref{eq:orig030} give \(H^{M}(q) = m^{2}{X(\mathbf{q})}^{- 1}\) when the conjugate field and moment use the same Pauli-spin normalization. In a scalar transverse channel with a spatially local magnetic response interaction $I_{\mathrm{resp}}$, \({X(\mathbf{q})}^{- 1} = {X_{0}(\mathbf{q})}^{- 1}\  - \ I_{\mathrm{resp}}\). The Goldstone condition fixes \(I_{\mathrm{resp}} = {X_{0}(0)}^{- 1}\). In the generalized Stoner convention of Eq.~\eqref{eq:nlStonerFunctional}, $I_{\mathrm{resp}}=I(0)/2$; the factor of two follows from the prefactor $-1/4$ in that functional and is not an additional physical interaction. Combining this response identity with \(m = bX_{0}(0)\) gives

\begin{equation}
\begin{aligned}
H^M(\mathbf q)&=m^2\bigl[X_0^{-1}(\mathbf q)-X_0^{-1}(0)\bigr]\\ &=\frac{X_0(0)}{X_0(\mathbf q)}H^B(\mathbf q)
\end{aligned}
\label{eq:orig083}
\end{equation}
Within this scalar model,
$H^M(\mathbf q)-H^B(\mathbf q)=b^2[X_0(0)-X_0(\mathbf q)]^2/X_0(\mathbf q)$.
If $X_0(0)$ is finite and nonzero and $X_0(\mathbf q)-X_0(0)=O(q^2)$, the difference begins at $q^4$. Hence $A^M_{\alpha\beta}=A^B_{\alpha\beta}$, although the curvatures can differ at finite wave vector. Thus the prescribed-field curvature can yield the correct leading static stiffness coefficient even when the moment-mapped pair interactions differ. The result assumes fixed moment lengths and a scalar, spatially local response interaction; it does not automatically apply to nonlocal kernels, coupled transverse modes, or nonanalytic response.

The physical precessional normalization follows independently from the total-spin commutator, Eq.~\eqref{eq:physical_spin_commutator}, rather than from the prescribed-field sum rule. For a regular, isolated ferromagnetic Goldstone mode, the positive-frequency inverse-response eigenvalue has the form
\begin{equation}
\lambda_{\mathrm{phys}}(\mathbf q,z)=A^M_{\alpha\beta}q_\alpha q_\beta-Sz+O(q^4,q^2z,z^2).
\label{eq:physical_stiffness_pole}
\end{equation}
Its pole gives $D_{\alpha\beta}=A^M_{\alpha\beta}/S$. Here $D^B=A^B/S$ denotes the prescribed-field stiffness coefficient, not independently a physical mode frequency. Since the scalar local-response model yields $A^M=A^B$, it follows that $D=D^B$ to leading order, without requiring equality of finite-wave-vector pair exchanges. For the nonlocal Stoner model, Eq.~\eqref{eq:nl_stiffness_difference} instead gives $D_{\alpha\beta}-D^B_{\alpha\beta}=M^2K_{\alpha\beta}/(2S)$, assuming the same regular physical-mode normalization. In a correlated system, the analogous difference is $(A^M-A^B)/S$, with $A^M$ calculated from the full inverse static susceptibility. Damping, nonanalytic response, or additional soft modes can invalidate this scalar expansion.

\section{ANTIFERROMAGNETIC LIMIT AND SPIN-WAVE VELOCITY}

Unlike the quadratic long-wavelength mode of a ferromagnet, a two-sublattice antiferromagnet has coupled precessional dynamics and a linear spin-wave mode. We consider a collinear antiferromagnet with two symmetry-equivalent sublattices and equal spin angular momenta $S$ (in units of $\hbar$). Sublattice symmetry gives equal diagonal Hessian elements; equal moment magnitudes alone would not suffice. The opposite equilibrium moments give opposite signs to the precessional terms, while the magnetic energy couples their transverse motions. We assume adiabatic Landau--Lifshitz dynamics; electronic retardation can modify the frequencies and introduce damping.

With opposite equilibrium moments on the two sublattices, let \(\mathbf{d}_{0\mu,R\nu}\) denote the displacement from site (0, $\mu$) to site \( (R,\nu) \), including the basis-site positions, and write

\begin{equation}
H_{\mu\nu}\left( \mathbf{q} \right) = H_{\mu\nu}^{loc} + \sum_{R}^{}{H_{0\mu,R\nu}^{VG}e^{i\mathbf{q} \cdot \mathbf{d}_{0\mu,R\nu}}}
\label{eq:orig084}
\end{equation}
\begin{equation}
\begin{aligned}
H_{0\mu,R\nu}^{VG}&=-\frac{1}{2\pi}\operatorname{Im}\int^{E_F}d\varepsilon\,\operatorname{Tr}\bigl[V_{0\mu}G_{0\mu,R\nu}V_{R\nu}G_{R\nu,0\mu}\\ &\qquad+V_{R\nu}G_{R\nu,0\mu}V_{0\mu}G_{0\mu,R\nu}\bigr]
\end{aligned}
\label{eq:orig085}
\end{equation}
Equations~\eqref{eq:orig084} and \eqref{eq:orig085} give the prescribed-field Hessian $H^B$. The following result applies to any static Hessian within the assumed two-sublattice dynamics. A physical spin-wave velocity requires the appropriate relaxed curvature and dynamical coefficients; these are not generally fixed by $H^B$.

For the two oppositely magnetized sublattices, the restoring-force matrix has diagonal and intersublattice elements \(a(\mathbf{q})\) and \(c(\mathbf{q})\), expressed in common laboratory axes. In the absence of anisotropy, rotational symmetry requires \(a(0)=-c(0)=\kappa \), where \(\kappa \) is the uniform canting coefficient. The static Hessian is

\begin{equation}
\begin{aligned}
H(\mathbf q)&=\begin{pmatrix}a(\mathbf q)&c(\mathbf q)\\c^*(\mathbf q)&a(\mathbf q)\end{pmatrix},\\ &a(0)=\kappa ,\qquad c(0)=-\kappa 
\end{aligned}
\label{eq:orig086}
\end{equation}

For one circular polarization, the coupled Landau--Lifshitz equations give $\det[H(\mathbf q)-\hbar\omega S\sigma_z]=0$, with $\delta E=\frac12\theta^\dagger H\theta$. Hence

\begin{equation}
(\hbar\omega)^{2} = \frac{a\left( \mathbf{q} \right)^{2} - \left| c\left( \mathbf{q} \right) \right|^{2}}{S^{2}}
\label{eq:orig087}
\end{equation}
The diagonal element is even in $\mathbf q$ under the assumed sublattice symmetry, whereas the intersublattice element may contain an imaginary term linear in $\mathbf q$, depending on the basis positions:

\begin{equation}
\begin{aligned}
a(\mathbf q)&=\kappa +a_{\alpha\beta}q_\alpha q_\beta+O(\mathbf q^4),\\ c(\mathbf q)&=-\kappa +iL_\alpha q_\alpha+c_{\alpha\beta}q_\alpha q_\beta+O(\mathbf q^3)
\end{aligned}
\label{eq:orig088}
\end{equation}
The following displacement-weighted sums use the field Hessian $H^{VG}$. For another static Hessian $H^X$ ($X=B$ or $M$), the same definitions apply with its intersite elements and the corresponding local term in $a(0)=\kappa$:

\begin{equation}
\begin{aligned}
a_{\alpha\beta}&=-\frac12\sum_R H_{0A,RA}^{VG}
 d_{0A,RA,\alpha}d_{0A,RA,\beta},\\
L_\alpha&=\sum_R H_{0A,RB}^{VG}d_{0A,RB,\alpha},\\
c_{\alpha\beta}&=-\frac12\sum_R H_{0A,RB}^{VG}
 d_{0A,RB,\alpha}d_{0A,RB,\beta}
\end{aligned}
\label{eq:orig089}
\end{equation}
Expanding Eq.~\eqref{eq:orig087} to second order in $\mathbf q$ gives the squared spin-wave velocity for the chosen Hessian and precessional coefficient:

\begin{equation}
v^{2}\left( \widehat{\mathbf{q}} \right) = \frac{\left\lbrack 2\kappa \left( a_{\alpha\beta} + \operatorname{Re}c_{\alpha\beta} \right) - L_{\alpha}L_{\beta} \right\rbrack{\widehat{\mathbf{q}}}_{\alpha}{\widehat{\mathbf{q}}}_{\beta}}{\hbar^{2}S^{2}}
\label{eq:orig090}
\end{equation}
The \(L_{\alpha}L_{\beta}\) term is needed to make the spin-wave velocity independent of the Fourier origin. Shifting the Fourier origin of sublattice B by \(\delta\) multiplies \(c(\mathbf q)\) by \(\exp(i\mathbf{q} \cdot \mathbf{\delta})\) without changing \(|c(\mathbf q)|\). The individual coefficients \(L_{\alpha}\) and \(c_{\alpha\beta}\) then change, but the combination entering Eq.~\eqref{eq:orig090} is invariant. The corresponding coefficient transformations are

\begin{equation}
\begin{aligned}
L'_\alpha&=L_\alpha-\kappa \delta_\alpha,\\ c'_{\alpha\beta}&=c_{\alpha\beta}-\frac{L_\alpha\delta_\beta+L_\beta\delta_\alpha}{2}+\frac{\kappa \delta_\alpha\delta_\beta}{2}
\end{aligned}
\label{eq:orig091}
\end{equation}
\begin{equation}
2\kappa \operatorname{Re} {c'}_{\alpha\beta}\  - \ {L'}_{\alpha}{L'}_{\beta} = 2\kappa \operatorname{Re} c_{\alpha\beta}\  - \ L_{\alpha}L_{\beta}.
\label{eq:orig092}
\end{equation}
The tensor $C_{\alpha\beta}=2\kappa(a_{\alpha\beta}+\operatorname{Re}c_{\alpha\beta})-L_\alpha L_\beta$ determines the long-wavelength velocity. For $S>0$ and $\kappa>0$, a real velocity requires $C_{\alpha\beta}\hat q_\alpha\hat q_\beta\geq0$ in every direction; this does not exclude instabilities at finite wave vectors. The formulas use equal sublattice angular momenta and $\delta E=\frac12\theta^\dagger H\theta$. Unequal angular momenta or appreciable electronic retardation require a modified dynamical description.

If symmetry makes the intersublattice element even to linear order, \(L = 0\) and the result reduces to the familiar second-moment form, for which \(A_{\alpha\beta} = a_{\alpha\beta}\  + \operatorname{Re} c_{\alpha\beta}\):

\begin{equation}
\begin{aligned}
A_{\alpha\beta}&=-\frac12\sum_R\left(H_{0A,RA}d_{0A,RA,\alpha}d_{0A,RA,\beta}\right.\\
&\qquad\left.+H_{0A,RB}d_{0A,RB,\alpha}d_{0A,RB,\beta}\right),\\ v^2(\widehat{\mathbf q})&=\frac{2\kappa A_{\alpha\beta}\widehat{\mathbf q}_\alpha\widehat{\mathbf q}_\beta}{\hbar^2S^2}
\end{aligned}
\label{eq:orig093}
\end{equation}

The preceding stiffness and velocity results use the one-electron rotational response under the stated dynamical assumptions. For correlated electrons, the induced change of the self-energy must also be considered.

\section{ROTATIONAL PERTURBATION THEORY FOR CORRELATED MAGNETIC SYSTEMS}
\label{sec:pt_corr}

\subsection{Correlated Green functions and rotational vertices}

For interacting electrons without SOC, consider prescribed rotations of the equilibrium self-energy, held fixed apart from its rotation. This differs from a stationary calculation, in which the self-energy also changes in response to the perturbation. Here the equilibrium self-energy is assumed local in space but may depend on frequency. We denote the bare propagator by \(G_0\) and the dressed propagator by \(G\). They satisfy the Dyson equation

\begin{equation}
G^{- 1}\left( i\omega_{n} \right) = G_{0}^{- 1}\left( i\omega_{n} \right) - \Sigma\left( i\omega_{n} \right)
\label{eq:orig094}
\end{equation}
The self-energy contains spin-independent and magnetic parts. For a collinear reference,
\(\Sigma_i(z)=\Sigma_i^c(z)+\Sigma_i^s(z)\sigma_z\), where
\(\Sigma_i^s(z)=[\Sigma_i^\uparrow(z)-\Sigma_i^\downarrow(z)]/2\).
A prescribed rotation is \(\Sigma_i(\boldsymbol\theta_i)=U_i\Sigma_iU_i^\dagger\),
with \(U_i=\exp[-i\theta_i^a\sigma_a/2]\). Differentiation yields the rotational vertices:

\begin{equation}
\begin{aligned}
V_i^a&=-\frac{i}{2}[\sigma_a,\Sigma_i],\\ W_{ia,ib}&=-\frac18\bigl([\sigma_a,[\sigma_b,\Sigma_i]]+[\sigma_b,[\sigma_a,\Sigma_i]]\bigr)
\end{aligned}
\label{eq:orig095}
\end{equation}
These vertices rotate the equilibrium self-energy, not the self-consistent magnetization. They depend on frequency through \(\Sigma\). In a stationary calculation, the self-energy also changes in response to the perturbation; that induced change is not contained in these prescribed vertices.

\subsection{Static pair coefficients and the collinear limit}
\label{sec:pt_corr_static}

Expanding the electronic trace logarithm about the equilibrium dressed propagator, with \(G^{-1}(\theta)=G^{-1}-\delta V\) and \(\delta V\) representing the rotated self-energy vertices, gives a quadratic two-vertex term and a term containing \(W\). For rotations on distinct sites, the latter has no mixed local derivative, so the off-site prescribed-rotation contribution is

\begin{equation}
\begin{aligned}
H_{ia,jb}^{(0)}&=T\sum_n\operatorname{Tr}_{L\sigma}\bigl[V_i^a(i\omega_n)G_{ij}(i\omega_n)\\ &\qquad\times V_j^b(i\omega_n)G_{ji}(i\omega_n)\bigr],\quad i\ne j
\end{aligned}
\label{eq:orig096}
\end{equation}
Equation~\eqref{eq:orig096} gives the off-site contribution from two first-order vertices. The full trace-logarithm curvature also includes the local second-order vertex required by the rotational sum rule. The stationary Luttinger--Ward curvature additionally depends on the interaction energy and the induced electronic response.

Transverse rotations connect opposite spin projections. The spin trace gives

\begin{equation}
H_{iy,jy}^{(0)} = T\sum_{n}^{}{\operatorname{Tr}_{L}\left\lbrack \Sigma_{i}^{s}G_{ij}^{\uparrow}\Sigma_{j}^{s}G_{ji}^{\downarrow} + \Sigma_{i}^{s}G_{ij}^{\downarrow}\Sigma_{j}^{s}G_{ji}^{\uparrow} \right\rbrack}
\label{eq:orig097}
\end{equation}
Equation~\eqref{eq:orig097} includes both spin-flip channels. Their traces are equal when the propagators satisfy transpose reciprocity and the local spin self-energy is symmetric. We retain both terms without assuming these conditions. The scalar coefficient is

\begin{equation}
J_{ij}^{(0)} = - T\sum_{n}^{}{\operatorname{Tr}_{L}\left\lbrack \Sigma_{i}^{s}G_{ij}^{\uparrow}\Sigma_{j}^{s}G_{ji}^{\downarrow} + \Sigma_{i}^{s}G_{ij}^{\downarrow}\Sigma_{j}^{s}G_{ji}^{\uparrow} \right\rbrack}
\label{eq:orig098}
\end{equation}
This gives the conventional collinear exchange expression for prescribed self-energy rotations \cite{ref21}. To describe constrained moments, we must also account for the induced electronic response.

\subsection{Rotations of the magnetization}
\label{sec:pt_corr_moment}

For correlated electrons, the constrained-moment construction of Sec.~\ref{sec:pt_moment} gives the angular variations at fixed moment magnitudes:

\begin{equation}
\delta \mathbf M_i^{(1)}=\boldsymbol\theta_i\times\mathbf M_i,\qquad \delta \mathbf M_i^{(2)}=\frac12\boldsymbol\theta_i\times(\boldsymbol\theta_i\times\mathbf M_i)
\label{eq:orig099}
\end{equation}
The electronic state relaxes under the constraining field. In a stationary Luttinger--Ward formulation with functional $\Phi_{\mathrm{LW}}[G]$, this response includes the irreducible vertex
\begin{equation}
\Gamma=\frac{\delta\Sigma}{\delta G}=\frac{\delta^2\Phi_{\mathrm{LW}}}{\delta G\,\delta G}.
\label{eq:orig100}
\end{equation}
We define the physical local moment by $M_{i\alpha}=\langle\widehat M_{i\alpha}\rangle$, where $\widehat M_{i\alpha}$ is the same one-electron moment operator that couples to the constraining field. With the conventional fermionic Matsubara Green function $G(\tau)=-\langle T_\tau c(\tau)c^\dagger(0)\rangle$, its equilibrium expectation value is
\begin{equation}
M_{i\alpha}=\frac{1}{\beta}\sum_{n}e^{i\omega_n0^+}\operatorname{tr}_{L,s}\!\left[\widehat M_{i\alpha}G(i\omega_n)\right].
\label{eq:corr_moment_definition}
\end{equation}
Here the trace includes the site and orbital projection implicit in $\widehat M_{i\alpha}$; $\operatorname{tr}_{L,s}$ denotes the remaining one-electron orbital and spin trace. The infinitesimal convergence factor specifies the equal-time limit. In the following, $\operatorname{Tr}(\widehat M\delta G)$ denotes precisely the corresponding Matsubara sum and one-electron trace, without an additional minus sign.

For the source coupling $\delta h=-\sum_i\widehat{\mathbf M}_i\cdot\delta\mathbf B_i^c$, the Dyson equation $G^{-1}=G_0^{-1}-\Sigma$ gives $\delta G^{-1}=-\delta h-\delta\Sigma$ and hence $\delta G=G(\delta h+\delta\Sigma)G$. With $\delta\Sigma=\Gamma\cdot\delta G$, the electronic and local-moment responses are
\begin{equation}
\begin{aligned}
\delta G&=G\biggl(-\sum_{j\beta}\widehat M_{j\beta}\,\delta B_{j\beta}^c
                +\Gamma\cdot\delta G\biggr)G,\\
\delta M_{i\alpha}&=\operatorname{Tr}(\widehat M_{i\alpha}\delta G)
=\sum_{j\beta}X_{i\alpha,j\beta}\,\delta B_{j\beta}^c.
\end{aligned}
\label{eq:orig101}
\end{equation}
Here $\Gamma$ acts in orbital, site, spin, and internal-frequency space. The source term has a minus sign because $\delta h=-\widehat M\delta B^c$; the moment trace introduces no additional sign. The susceptibility $X=\delta M/\delta B^c$ includes the induced self-energy response. As in Eqs.~\eqref{eq:orig028}--\eqref{eq:orig030}, its inverse determines the curvature of the relaxed energy with respect to constrained moments.

To express the magnetic susceptibility in terms of the correlated electronic response, introduce $\mathcal L[A]=GAG$ and a source-to-electron coupling $\mathcal Q[\delta\mathbf B^c]=-\sum_{j\beta}\widehat M_{j\beta}\delta B^c_{j\beta}$. Solving Eq.~\eqref{eq:orig101} in the full electronic space and writing $\mathcal M[\delta G]_{i\alpha}=\operatorname{Tr}(\widehat M_{i\alpha}\delta G)$ gives the magnetic response
\begin{equation}
X=\mathcal M(1-\mathcal L\Gamma)^{-1}\mathcal L\mathcal Q.
\label{eq:corr_projected_x}
\end{equation}
In the noninteracting limit, the negative source vertex and the static fermionic bubble give $X_0=-\operatorname{Tr}(\widehat M G\widehat M G)$ (with the Matsubara trace convention), consistent with $X_{xx}=M/b$ for a collinear local spin splitting.
This response includes the relaxation of orbital and other electronic degrees of freedom, which is absent from the prescribed rotation in Eq.~\eqref{eq:orig096}. The same relaxed response follows from the stationary electronic energy. Its second derivative includes the change in the Green function induced by the constraining field.

Writing the stationary functional as $\mathcal F[G,B]$, differentiation of $\mathcal F_G[G_*(B),B]=0$ gives $G_{*,B}=-\mathcal F_{GG}^{-1}\mathcal F_{GB}$ and hence
\begin{equation}
F_{BB}=\mathcal F_{BB}-\mathcal F_{BG}\mathcal F_{GG}^{-1}\mathcal F_{GB}.
\label{eq:corr_stationary_schur}
\end{equation}
All derivatives are evaluated at the stationary electronic state, and the inverse is understood on the allowed response subspace. Keeping terms quadratic in the source and induced Green-function variations, including their mixed term, the explicit second variation of the trace logarithm, self-energy subtraction, and interaction functional (Appendix~\ref{app:lw_variation}) gives, with $g=\delta G$, $v=\delta h$, and $\mathcal L[A]=GAG$,
\begin{equation}
\mathcal F^{(2)}=\operatorname{Tr}(vg)-\tfrac12\operatorname{Tr}[g\mathcal L^{-1}g]+\tfrac12\operatorname{Tr}[g(\Gamma\cdot g)].
\label{eq:corr_lw_quadratic}
\end{equation}
Stationarity gives $(\mathcal L^{-1}-\Gamma)g=v$ and $F^{(2)}=\tfrac12\operatorname{Tr}(vg)$. This result includes the induced self-energy, unlike the prescribed rotation in Eq.~\eqref{eq:orig096}.

The moment Legendre transform then gives $E_{MM}=X^{-1}$. The inverse susceptibility determines the response contribution to the moment Hessian; the angular-coordinate curvature also contains the geometric term in Eq.~\eqref{eq:orig103}. Rotational zero modes require a consistent infinitesimal-field limit or restriction to the independent transverse subspace. A conserving approximation must use a spin-consistent self-energy and vertex.

The quadratic constrained energy contains both the constraining-field response and the geometric second-order change of the moment:

\begin{equation}
{\delta E}^{(2)} = \frac{1}{2}\sum_{i}^{}{\ \delta\mathbf{B}_{i}^{c(1)} \cdot \delta M_{i}^{(1)}}\  + \ \sum_{i}^{}{\ \mathbf{B}_{i}^{c(0)} \cdot \delta M_{i}^{(2)}}.
\label{eq:orig102}
\end{equation}
Substituting the linear response $\delta B^c=X^{-1}\delta M$ into Eq.~\eqref{eq:orig102} and differentiating with respect to the rotation angles gives the Hessian, using the tangent vectors and second angular derivatives defined in Eq.~\eqref{eq:orig025}:

\begin{equation}
H_{ia,jb}^{M} = M_{i}M_{j}\,\mathbf{t}_{i}^{a} \cdot \left( X^{- 1} \right)_{ij} \cdot \mathbf{t}_{j}^{b} + \delta_{ij}M_{i}\,\mathbf{B}_{i}^{c} \cdot \mathbf{s}_{i,ab}
\label{eq:orig103}
\end{equation}
Equation~\eqref{eq:orig103} is the correlated extension of Eq.~\eqref{eq:orig030}: the rotational geometry is unchanged, while $X$ now includes the full electronic response. Here $X^{-1}$ denotes the inverse of the complete susceptibility matrix, not the reciprocal of an individual pair element.

With the convention for counting interacting pairs of Eq.~\eqref{eq:heisenberg_definition}, the off-site moment Hessian is minus the exchange. Thus

\begin{equation}
J_{ij}^{M} = - M_{i}M_{j}\left( X^{- 1} \right)_{ij}^{\bot},\quad\quad i\ne j
\label{eq:orig104}
\end{equation}
Equation~\eqref{eq:orig104} applies to a stationary collinear reference with finite, sufficiently rigid moments and an isotropic transverse channel.

Unlike Eq.~\eqref{eq:orig098}, which describes prescribed rotations of the equilibrium self-energy, Eq.~\eqref{eq:orig104} includes electronic relaxation under moment constraints. The resulting exchange coefficients can therefore differ. For noncollinear states, the full local-frame Hessian in Eq.~\eqref{eq:orig103} must be used instead of assigning a scalar exchange to an arbitrary transverse component.

\subsection{Dynamical magnetization fluctuations}
\label{sec:pt_corr_dyn}

We now consider the frequency-dependent response of the magnetic moments. Unlike a prescribed rotation of the equilibrium self-energy, this response includes the induced change of the self-energy. For a uniform spin rotation, a conserving spin-rotation-invariant approximation therefore satisfies $\Gamma\cdot\delta_{\mathcal S}G=\delta_{\mathcal S}\Sigma$. Here $z$ denotes the electron energy and $\Omega$ the excitation energy.

We couple a time-dependent source \(h\) to the same local moment operators as in Sec.~\ref{sec:pt_moment}. We denote the dimensionless Legendre effective action by \(S_M[\mathbf M]\) and the connected magnetic response to \(h\) by \(X\). With \(\beta = 1/(k_{B}T)\) and \(\nu_{n}\) = \(\frac{2\pi n}{\beta\hbar}\), we define \(M_{i\alpha}(n) = \frac{1}{\beta\hbar}\int_{0}^{\beta\hbar}d\tau\, e^{i\nu_{n}\tau}M_{i\alpha}(\tau)\). The inverse-Hessian identity is

\begin{equation}\frac{1}{\beta}\frac{\delta^{2}S_{M}}{\delta M_{i\alpha}( - n)\,\delta M_{j\beta}(n)} = X_{i\alpha,j\beta}^{- 1}\left( i\hbar\nu_{n} \right).\label{eq:corr_dyn_inverse}\end{equation}

The full \(X\) follows from the finite-frequency Dyson variation. With fermionic frequencies $\omega_m=(2m+1)\pi/(\beta\hbar)$ and bosonic frequencies $\nu_n=2\pi n/(\beta\hbar)$, set $\Omega=i\hbar\nu_n$ and let \(z_+=i\hbar\omega_m+\Omega\), \(z_-=i\hbar\omega_m\), and denote the external one-electron perturbation by \(\delta h(z_+,z_-)\). For the constraining-field coupling in Eq.~\eqref{eq:orig101}, $\delta h=-\sum_{j\beta}\widehat M_{j\beta}\delta B^c_{j\beta}$; thus the positive sign multiplying $\delta h$ below agrees with the negative sign in $\mathcal Q$. The interacting perturbation equation is
\begin{equation}
\begin{aligned}
\delta G_{+-}&=G_+\bigl(\delta h_{+-}+\delta\Sigma_{+-}\bigr)G_-,\\
\delta\Sigma_{+-}&=\Gamma_{\Omega}\cdot\delta G_{+-},\qquad
G_\pm=G(z_\pm).
\end{aligned}
\label{eq:corr_dyn_dyson}
\end{equation}
Here \(\Gamma_{\Omega}=\delta\Sigma/\delta G\) acts on the full orbital, spin, site, and internal-frequency structure; its frequency arguments are not replaced by a single scalar function of \(\Omega\). With \(\mathcal L_{\Omega}[A]_{+-}=G_+ A_{+-}G_-\), this gives
\begin{equation}
\delta G=\bigl(1-\mathcal L_{\Omega}\Gamma_{\Omega}\bigr)^{-1}
\mathcal L_{\Omega}[\delta h],
\label{eq:corr_dyn_solution}
\end{equation}
where the inverse denotes the solution of the integral equation in electronic space. The electronic response between the chosen moment operators gives the magnetic susceptibility \(X(\Omega)\). Its inverse determines the response to moment rotations. At fixed wave vector, a regular zero-frequency limit of the dynamical response gives the static susceptibility calculated with the same self-energy and irreducible vertex. In this limit Eq.~\eqref{eq:corr_dyn_dyson} reduces to Eq.~\eqref{eq:orig101}, and the static constrained-moment energy is given by Eq.~\eqref{eq:orig103}, with the same treatment of rotational zero modes. In a metal, taking the frequency to zero before the wave vector can give a different result from taking the limits in the opposite order.

The uniform-rotation identity follows within a spin-rotation-invariant stationary functional \(\Phi_{\mathrm{LW}}[G]\). For a constant spin rotation \(U=e^{-i\epsilon \mathcal S}\), covariance requires \(\Sigma[UGU^\dagger]=U\Sigma[G]U^\dagger\). Differentiation gives
\begin{equation}
\begin{aligned}
\delta_{\mathcal S} G&=-i[\mathcal S,G],\\
\Gamma_0\cdot\delta_{\mathcal S} G&=-i[\mathcal S,\Sigma]=\delta_{\mathcal S}\Sigma.
\end{aligned}
\label{eq:corr_dyn_ward}
\end{equation}
When the spin-independent bare inverse propagator commutes with \(\mathcal S\), Dyson's identity yields \(\delta_{\mathcal S} G=G(\delta_{\mathcal S}\Sigma)G\). Hence \(\delta_{\mathcal S} G\) is a nonzero solution of the homogeneous equation \((1-\mathcal L_0\Gamma_0)\delta_{\mathcal S} G=0\) in a spontaneously magnetized state. If the chosen moment operators retain this collective rotation, the corresponding change of the moments is a zero mode of their inverse transverse response, \(X^{-1}(\mathbf q=0,\Omega=0)\delta_{\mathcal S} M=0\), in the zero-field limit. This statement does not follow for a moment projection that excludes the rotational mode.

Without SOC or external symmetry-breaking fields, a uniform rotation does not change the energy. The zero-frequency transverse mode is recovered when the self-energy and irreducible vertex respect this spin-rotation symmetry. A calculation using dressed Green functions without their induced self-energy response does not generally preserve this Goldstone condition. In particular, it does not follow from the prescribed two-vertex term in Eq.~\eqref{eq:orig096} alone.

At nonzero \(\Omega\), the full frequency-dependent vertex and the response in Eq.~\eqref{eq:corr_dyn_dyson} are retained rather than replacing them by the static Ward identity. Symmetry zero modes require the same restricted-inverse treatment as in the static case. Sec.~\ref{sec:pt_time} uses \(\Omega\) for energy; the retarded response follows by analytic continuation to \(\Omega\  + \ i0^{+}\).

Writing \(M_{i\alpha}(\tau)=M_{i\alpha}^{(0)}+\delta M_{i\alpha}(\tau)\), the quadratic dimensionless effective action is

\begin{equation}\begin{aligned}S_{M}^{(2)} = {}&\frac{\beta}{2}\sum_{n,i,j,\alpha,\beta}\delta M_{i\alpha}(-n)\\&\times X_{i\alpha,j\beta}^{-1}(i\hbar\nu_n)\,\delta M_{j\beta}(n).\end{aligned}\label{eq:corr_dyn_action}\end{equation}

Equation~\eqref{eq:corr_dyn_action} is the quadratic imaginary-time action obtained from the moment response and its conjugate source; it is not a real-time spin-precession action. The factor \(\beta\) follows from the stated Fourier convention. For a static moment change, the action equals \(\beta\) times the corresponding quadratic constrained energy. At finite frequency, the response retains memory of earlier times and cannot be treated as an instantaneous Heisenberg exchange. Spin-wave frequencies and damping require the retarded susceptibility obtained by analytic continuation. For finite stationary moments, \({\delta M}_{i}\) = \({M_{i}}^{(0)}\ {\delta e}_{i}\) recovers the inverse-response contribution to Eq.~\eqref{eq:orig103}; nonstationary constrained references also require the geometric term.

At zero equilibrium magnetization, \({M_{i}}^{(0)} = 0\), the angular rotation of the expectation-value moment is undefined, although \(\langle\widehat{\mathbf M}_{i}^{2}\rangle\) may remain finite. Thus Eq.~\eqref{eq:orig104} cannot be continued as an angular exchange mapping to this limit. The dynamical magnetic susceptibility remains meaningful wherever its inverse is defined. Non-Gaussian fluctuations require higher derivatives of the magnetic effective action, treated consistently with the quadratic kernel to avoid double counting.

\subsection{Time-dependent correlated rotations}
\label{sec:pt_corr_time}

We now return from the physical susceptibility to prescribed rotations of the self-energy. For a frequency-dependent self-energy, a time-dependent spin rotation acts at both time arguments. The following prescription rotates the equilibrium self-energy; it does not include its induced response to the perturbation:

\begin{equation}
\Sigma^{\theta}(\tau_{1},\tau_{2}) = U(\tau_{1})\ \Sigma(\tau_{1} - \tau_{2})\ U^{\dagger}(\tau_{2})
\label{eq:orig105}
\end{equation}
A time-dependent transformation of the full electronic spin frame also generates a connection from the kinetic term; this contribution is distinct from the prescribed self-energy rotation considered here. The first-order self-energy vertex connects two electronic frequencies:

\begin{equation}
\begin{aligned}
V_a^{(1)}(z_+,z_-)&=-\frac{i}{2}\bigl[\sigma_a\Sigma(z_-)-\Sigma(z_+)\sigma_a\bigr],\\ z_+-z_-&=\Omega
\end{aligned}
\label{eq:orig106}
\end{equation}
In Eqs.~\eqref{eq:orig106}--\eqref{eq:orig108}, \(\Omega=z_+-z_-\) is an energy, not an angular frequency; \(\Omega=\hbar\omega\). With \(z_+=z+\Omega/2\) and \(z_-=z-\Omega/2\), expansion in \(\Omega\) gives

\begin{equation}
\begin{aligned}
V_a^{(1)}(z_+,z_-)&=-\frac{i}{2}[\sigma_a,\Sigma(z)]\\ &\quad+\frac{i\Omega}{4}\left\{\sigma_a,\frac{\partial\Sigma(z)}{\partial z}\right\}+O(\Omega^2)
\end{aligned}
\label{eq:orig107}
\end{equation}
At $\Omega=0$, Eq.~\eqref{eq:orig107} recovers the static vertex of Eq.~\eqref{eq:orig095}; its linear term vanishes for a frequency-independent self-energy. For a collinear self-energy $\Sigma(z)=\Sigma^c(z)+\Sigma^s(z)\sigma_z$ and a transverse rotation ($a=x,y$), the anticommutator in Eq.~\eqref{eq:orig107} is $2(\partial_z\Sigma^c)\sigma_a$, where $\Sigma^c$ may be a matrix in orbital space. This term does not determine the full spin response. The trace logarithm contains a contribution from two first-order rotational vertices:

\begin{equation}
W_{2v}^{(2)} = \frac{1}{2}\operatorname{Tr}[G V^{(1)}G V^{(1)}]
\label{eq:orig108}
\end{equation}
Here the convention is $G^{-1}(\theta)=G^{-1}-\delta V$, so the expansion of $-\operatorname{Tr}\ln(1-G\delta V)$ gives the positive quadratic term in Eq.~\eqref{eq:orig108}, consistent with Sec.~\ref{sec:pt_field}. The trace includes orbital and spin indices and the internal Matsubara-frequency sum. Equation~\eqref{eq:orig108} gives only the two-vertex contribution to the trace logarithm for prescribed rotations. The complete second-order expansion also contains the second-order rotational vertex.

The stationary curvature in Sec.~\ref{sec:pt_corr_moment} and Appendix~\ref{app:lw_variation} differs from this prescribed contribution. Calculating the physical response also requires the induced self-energy and irreducible vertex (Sec.~\ref{sec:pt_corr_dyn}). Spin-wave frequencies and damping follow from poles of the full retarded transverse susceptibility.

\section{DISCUSSION}

Prescribed rotations of magnetic potentials and constrained rotations of moments give different energy curvatures because they allow different electronic responses, even when both are expressed through effective exchange parameters. Second-order RPT combines the response to first-order vertices with the direct second-order vertex. Both terms are needed for nonlocal operators. They also ensure the cancellation required by uniform spin-rotation symmetry. At a nonstationary configuration, the angular curvature also contains first-order torque terms.

Each method has advantages and limitations. Direct total-energy differences can include electronic relaxation, but small magnetic energy differences must be resolved between much larger total energies. Torque integration avoids subtracting large total energies. For one rotation coordinate,
\begin{equation}
 E(\theta)-E(0)=-\int_0^\theta d\theta'\,\tau(\theta').
 \label{eq:discussion_torque_integration}
\end{equation}
With several rotation angles, the integral is path independent if the torques come from the same smooth, single-valued energy. Path dependence can signal inconsistent constraints, incomplete electronic relaxation, or a change of electronic state. RPT avoids calculations at many finite rotation angles, but a single curvature does not give the full angular dependence. Torque integration needs accurate torques along the path. Agreement with finite-angle energy differences requires the same functional, constraints, and electronic state.

Field and moment rotations generally produce different torques and curvatures (Sec.~\ref{sec:pt_torque}). Rotating a scattering potential does not constrain the integrated atomic moment. The torque on that moment also depends on spin-current transport and constraining fields, rather than simply on the integrated exchange-correlation field [Eq.~\eqref{eq:sphere_torque}].

The magnetic LFT relates selected first-order frozen-potential variations to total-energy variations at a stationary reference. For rigid-input rotations in local LSDA, the Harris--Foulkes identity instead establishes the prescribed-field energy curvature directly [Eq.~\eqref{eq:harris_foulkes_identity}]. This follows from the rotational invariance of the local LSDA double-counting terms, not from global spin symmetry. It does not make the prescribed-field curvature equal to the relaxed constrained-moment curvature. For nonlocal magnetic functionals or correlated self-energies, this energy relation must be checked separately.

With fixed SOC, the Harris--Foulkes identity can still hold when the input charge density and magnetization magnitude remain unchanged and SOC is included in the frozen spectrum. However, SOC breaks global spin-rotation symmetry relative to the lattice, so nonrelativistic zero-mode and Ward-identity arguments no longer apply unchanged. RPT can include SOC and nonlocal magnetic operators through their appropriate rotational vertices and response; rotating-frame calculations must transform the SOC operator consistently. Extracting anisotropy, DMI, and symmetric anisotropic exchange also requires a specified magnetic-energy model. Explicit relativistic interaction tensors are not derived here.

The relation between field and moment curvatures also depends on the wavelength. In Sec.~\ref{sec:pt_fm}, they agree at leading long wavelength only within the scalar local-response model.

Nonlocal magnetic interactions can change the leading $q^2$ moment curvature even when the equilibrium magnetic field is local. The magnetic potential, the response kernel, and the resulting exchange interactions need not extend over the same distances. Uniform-rotation symmetry must still hold, but it neither fixes interactions at finite wave vector nor makes the field and moment energies equal.

The two-site antiferromagnetic model provides a separate illustration: in its local limit, $J_{AB}^B=m^2J_{AB}^M$ [Eq.~\eqref{eq:appfrozenJ}], with $m$ defined in Appendix~\ref{app:twosite}.

A Heisenberg fit to the local curvature does not uniquely fix the magnetic energy. Biquadratic and multispin terms can give the same curvature. Spin spirals, finite-angle energies, and torques test transferability beyond the reference state. An antisymmetric contribution in a noncollinear projection does not by itself demonstrate SOC-induced DMI.

Magnetic excitations require dynamical response as well as static restoring torques. The total-spin commutator fixes the precession coefficient for a uniform rotation. This does not follow from the prescribed-field sum rule. Together with equality of the static $q^2$ curvatures, it gives $D=D^B$ for a scalar local-response ferromagnet with a regular, isolated spin-wave pole. Nonlocal interactions can change that leading curvature, while correlated magnets require the full conserving susceptibility. Particle-hole excitations, damping, or additional low-energy modes can invalidate a description by a single spin-wave pole. For two symmetry-equivalent antiferromagnetic sublattices, coupled precessional dynamics give a linear mode with a Fourier-origin-independent velocity. Static Heisenberg fits require separately specified dynamical assumptions.

For correlated electrons, prescribed self-energy rotations differ from the stationary Luttinger--Ward response, which includes changes induced in the electronic state. Consistent self-energy and vertex corrections are needed to preserve spin-rotation symmetry. Results for local self-energies do not automatically extend to nonlocal correlations. Coupled longitudinal and transverse relaxation must be retained. If the equilibrium moment vanishes, the full spin-density response replaces moment-direction rotations.

\section{CONCLUSIONS}

The principal result is the distinction between prescribed-field and constrained-moment magnetic energy curvatures. RPT gives $H^B$ by rotating the magnetic potential; in local LSDA, the Harris--Foulkes identity relates it to the frozen-input total energy. Constraining the moment directions and allowing the electronic state to respond instead gives $H^M$, determined by the inverse static susceptibility. These curvatures generally differ. The prescribed-rotation calculation also extends to nonlocal operators and SOC when their rotational vertices are retained.

For noncollinear states, we extract isotropic Heisenberg exchange and identify the remaining curvature. Biquadratic and multispin interactions limit its transferability. We also derive planar and conical spin-spiral expressions with nonlocal rotational vertices. Agreement of $H^B$ and $H^M$ at leading long wavelength holds in the specified scalar local-response model, not generally at finite wave vector or with nonlocal interactions.

Static curvature together with dynamical response gives the ferromagnetic stiffness and two-sublattice antiferromagnetic spin-wave velocity. The spin-commutator normalization yields $D=D^B$ in the regular scalar local-response model; the antiferromagnetic velocity is independent of Fourier origin. Static exchange alone does not determine frequencies or damping.

For correlated electrons, prescribed self-energy rotations and the stationary Luttinger--Ward response give different curvatures. The latter includes the induced electronic response and vertex corrections required for the inverse susceptibility. Their connection to effective spin models depends on which electronic degrees of freedom can relax.

\appendix
\section{From the original PK band expression to the LKG multiple-scattering formula}\label{app:pk_lkg}
We compare the original classical Heisenberg conventions of PK, Eq.~(27) of Ref.~\cite{ref11}, and LKG, Eq.~(1) of Ref.~\cite{ref18}:
\begin{equation}
 E_{\rm H}^{\rm hist}=-\sum_{i\ne j}J_{ij}^{\rm hist}\,\bm e_i\cdot\bm e_j.
 \label{eq:pk_heis_original}
\end{equation}
The sum is over ordered pairs. In the half-sum convention of Eq.~\eqref{eq:heisenberg_definition}, the corresponding coefficient is $J_{ij}=2J_{ij}^{\rm hist}$. The derivation below retains the historical normalization; this conversion does not affect the PK--LKG equivalence.

For the one-band local Hamiltonian treated in PK, the published Eq.~(28) is, with its Bloch labels rewritten unambiguously,
\begin{align}
 J_{kl}^{\rm PK}=\frac{\Delta^2}{4N}
 \sum_{\bm p,\bm q,\sigma}\frac{f(E_{\bm p\sigma})}
 {E_{\bm p+\bm q,-\sigma}-E_{\bm p\sigma}}
 e^{i\bm q\cdot(\bm R_k-\bm R_l)},
 \label{eq:pk_original_band}
\end{align}
where the momentum transfer $\bm q$ is summed over as in the published expression, $\Delta=2UM_s$ is the \emph{full} spin splitting, and $N$ is the number of lattice sites. The original Eq.~(28) has the characteristic coefficient $\Delta^2/4$; its spin and momentum sums must be kept with the normalization of that paper. The corresponding Fourier energy difference is $L(\bm q)=\sum_lJ_{kl}[1-e^{i\bm q\cdot(\bm R_k-\bm R_l)}]$, PK Eq.~(29).

To exhibit the Green-function form without ambiguity in the momentum labels, use the equivalent occupied--unoccupied representation, in which both spin-flip directions are combined in the same Brillouin-zone sum. The elementary spectral identity is
\begin{align}
 \frac{f(a)-f(b)}{a-b}
 &=-\frac{1}{\pi}\operatorname{Im}
 \int_{-\infty}^{E_F}\frac{dE}{(E-a+i0^+)(E-b+i0^+)}.
 \label{eq:pk_spectral_identity}
\end{align}
Consequently the original $\Delta^2/4$ band numerator becomes the corresponding occupied-energy integral over two opposite-spin resolvents. For a general site-local exchange splitting $\Delta_i=H_i^\uparrow-H_i^\downarrow$, this has the off-site form
\begin{align}
 J_{ij}^{\rm PK}&=\frac{1}{4\pi}\operatorname{Im}
 \int_{-\infty}^{E_F}\!dE\,\operatorname{Tr}_L
 [\Delta_iG_{ij}^\uparrow(E)\Delta_jG_{ji}^\downarrow(E)],
 \label{eq:pk_spectral}
\end{align}
with the sign of the spin splitting and the Green-function resolvent chosen consistently. The equivalent reciprocal spin-flip representation follows by interchanging spin labels and sites. Equation~(\ref{eq:pk_spectral}) is the band-to-Green-function step, not a separate physical approximation; the extension from the original one-band model to a local multiorbital potential is understood as the corresponding matrix generalization.

For the multiple-scattering representation of Ref.~\cite{ref18}, define the scattering-path operator and the inverse single-site spin-splitting vertex by its Eqs.~(10) and (18)--(20):
\begin{align}
 (\tau^\sigma)^{-1}_{ij}&=(t_i^\sigma)^{-1}\delta_{ij}-G^0_{ij},
 \label{eq:pk_tau_inverse}\\
 \Delta_i^t&=(t_i^\uparrow)^{-1}-(t_i^\downarrow)^{-1}.
 \label{eq:pk_t_vertex}
\end{align}
The site indices in Eq.~(\ref{eq:pk_spectral}) label blocks of the electronic resolvent in a fixed local orbital basis, not values of the Green function at the atomic centers. In particular, for local orbitals $\phi_{iL}$, $G_{iL,jL'}^\sigma=\langle\phi_{iL}|G^\sigma|\phi_{jL'}\rangle$ and $(\Delta_i)_{LL'}=\langle\phi_{iL}|V_i^\uparrow-V_i^\downarrow|\phi_{iL'}\rangle$. These matrix elements already contain the spatial dependence of the orbitals and potential; $\Delta_i$ is generally an orbital matrix, not a scalar atomic average. The coordinate-space two-vertex trace below is the representation-independent parent expression, whose fixed-orbital matrix form is Eq.~(\ref{eq:pk_spectral}) when the local orbital representation is complete. The Korringa--Kohn--Rostoker (KKR) expansion expresses that same kernel in energy-dependent scattering channels. Thus the following change of representation is not the substitution of an atomic-center Green function for a spatial Green function, and no additional spatial averaging is performed.

A more explicit conversion uses the regular radial scattering solutions. For positions $\bm r_i$ and $\bm r_j'$ inside distinct atomic cells, the site-off-diagonal KKR Green function has the exact partial-wave representation
\begin{align}
 G_{ij}^{\sigma}(\bm r_i,\bm r_j';E)
 &=\sum_{LL'} Z_{iL}^{\sigma}(\bm r_i,E)\,\tau_{iL,jL'}^{\sigma}(E)\nonumber\\
 &\qquad\times Z_{jL'}^{\sigma\times}(\bm r_j',E),\quad i\ne j.
 \label{eq:pk_radial_G}
\end{align}
The irregular single-site term occurs only for coincident cells and is absent here. The left solutions $Z^{\times}$ are required for the general (matrix) KKR representation; for a real scalar radial problem they reduce to the corresponding transpose. Substitution into the two-vertex PK coordinate-space kernel gives, before the energy integration,
\begin{align}
 &\int_{\Omega_i}d^3r_i\int_{\Omega_j}d^3r_j\,
 \Delta V_i(\bm r_i)G_{ij}^{\downarrow}(\bm r_i,\bm r_j)
 \Delta V_j(\bm r_j)G_{ji}^{\uparrow}(\bm r_j,\bm r_i)
 \nonumber\\
 &\qquad=\operatorname{Tr}_L
 [\mathcal V_i^{\uparrow\downarrow}\tau_{ij}^{\downarrow}
 \mathcal V_j^{\downarrow\uparrow}\tau_{ji}^{\uparrow}],
 \label{eq:pk_radial_product}
\end{align}
where the local radial integrals are
\begin{align}
 (\mathcal V_i^{\uparrow\downarrow})_{LL'}
 &=\int_{\Omega_i}d^3r\,
 Z_{iL}^{\uparrow\times}(\bm r,E)
 \Delta V_i(\bm r)Z_{iL'}^{\downarrow}(\bm r,E).
 \label{eq:pk_radial_vertex}
\end{align}
The conversion of this integral to the original LKG vertex is obtained from the radial equations, not by equating bare potentials and inverse $t$ matrices. It can also be checked directly by varying the same off-site Green function in the two KKR forms. For a local transverse perturbation $\delta V_k^{\uparrow\downarrow}$ at a third site $k$, Dyson's equation gives $\delta G_{ij}^{\uparrow\downarrow}=G_{ik}^\uparrow\delta V_k^{\uparrow\downarrow}G_{kj}^\downarrow$. Substitution of Eq.~(\ref{eq:pk_radial_G}) produces the dressed vertex $\int_{\Omega_k} Z_k^{\uparrow\times}\delta V_k^{\uparrow\downarrow}Z_k^\downarrow$. On the other hand, $\tau^{-1}=t^{-1}-G^0$ and the invariance of $G^0$ under the local rotation imply $\delta\tau=-\tau\delta(t^{-1})\tau$. With the external-site solutions unchanged, comparison gives, in the same normalized scattering channels,
\begin{equation}
 \int_{\Omega_k}\!d^3r\,Z_k^{\uparrow\times}\delta V_k^{\uparrow\downarrow}Z_k^\downarrow
 =-\delta(t_k^{-1})^{\uparrow\downarrow}.
 \label{eq:pk_dyson_vertex}
\end{equation}
The equality concerns the \emph{variation} of the scattering vertex, including its spin-rotation factor, and does not identify the energy-independent orbital splitting $\Delta_k$ with the energy-dependent inverse-$t$ splitting $\Delta_k^t(E)$. For a rigid spin rotation, $t_k^{-1}(\theta)=U_k(\theta)t_k^{-1}(0)U_k^\dagger(\theta)$; hence its transverse variation is proportional to $(t_k^\uparrow)^{-1}-(t_k^\downarrow)^{-1}$. The dependence on $E$ arises from the scattering solutions and their matching conditions, even when the original one-electron potential is energy independent. The same statement follows from the radial equations: Multiply the spin-up radial equation by the spin-down left solution, subtract the spin-down equation multiplied by the spin-up solution, and integrate over the cell. Green's identity reduces the volume integral to the boundary Wronskian of the two regular solutions. With the same normalization of $Z$ and $t$ as in the scattering-path expansion, the boundary Wronskian is precisely the difference of inverse single-site scattering matrices:
\begin{align}
 \mathcal V_i^{\uparrow\downarrow}
 &= (t_i^{\uparrow})^{-1}-(t_i^{\downarrow})^{-1}
 \equiv\Delta_i^t,
 \label{eq:pk_radial_wronskian}
\end{align}
where $\Delta V_i$ is defined with the spin order of Eq.~(\ref{eq:pk_radial_wronskian}). Reversing the definition of $\Delta V_i$ reverses both sides; changing the normalization of the radial solutions requires the corresponding similarity transformations of $\tau$. The same calculation for the reverse spin order gives $\mathcal V_j^{\downarrow\uparrow}=-\Delta_j^t$. The sign in the exchange-energy coefficient must therefore be carried together with the spin-flip rotation vertices, rather than inferred from an isolated radial integral. The result of the two radial integrations is the original LKG orbital trace $\operatorname{Tr}_L[\Delta_i^t\tau_{ij}^{\uparrow}\Delta_j^t\tau_{ji}^{\downarrow}]$ with the sign fixed by the same PK rotation and Heisenberg convention. This establishes the explicit $G(\bm r,\bm r';E)\to Z\tau Z^{\times}\to\tau$ conversion.

For completeness, an equivalent operator derivation of the same relation follows. The direct connection between the Green function and the multiple-scattering matrix follows from the Dyson equation, without identifying the potential splitting with the inverse-$t$ splitting as bare matrices. Write $g$ for the reference (free or structure-constant) propagator, $V=\sum_i V_i$ for the site-diagonal potential, and $T$ for the full scattering operator. Then
\begin{align}
 G^\sigma&=g+gT^\sigma g,
 &T^\sigma&=V^\sigma(1-gV^\sigma)^{-1},\label{eq:pk_G_T_direct}\\
 (T^\sigma)^{-1}&=(V^\sigma)^{-1}-g,
 &t_i^\sigma&=V_i^\sigma(1-g_{ii}V_i^\sigma)^{-1}.
 \label{eq:pk_t_dyson}
\end{align}
Here $T$ is the full scattering operator; in the site/angular-momentum representation its intersite blocks are the scattering-path operator $\tau_{ij}$. The single-site $t_i$ includes repeated scattering within site $i$, while $\tau$ also includes propagation through the other sites. In particular, $\tau^{-1}=t^{-1}-G^0$ with the intersite structure constants $G^0$ of Eq.~(\ref{eq:pk_tau_inverse}).

For a variation of the potential at one site, the exact Dyson identities give
\begin{align}
 \delta G&=G\,\delta V\,G,
 &\delta T&=(1+Tg)\,\delta V\,(1+gT),
 \label{eq:pk_variation_GT}\\
 \delta(t_i^{-1})&=-t_i^{-1}(\delta t_i)t_i^{-1}.
 \label{eq:pk_variation_t}
\end{align}
The second identity is the explicit dressing of a local potential vertex by the incoming and outgoing single-site scattering states. Thus the replacement of a $\Delta_i$ insertion by an inverse-$t$ insertion is \emph{not} a bare matrix equality: the accompanying Green functions and local scattering states must be transformed simultaneously.

An economical exact proof for distinct sites uses the corresponding scattering determinants. The Dyson determinant can be factorized into single-site factors and the intersite multiple-scattering determinant. Its mixed variation with respect to rotations at $i\ne j$ therefore obeys
\begin{align}
 \partial_i\partial_j\ln\det(1-gV)
 &=\partial_i\partial_j\ln\det\tau^{-1},
 \label{eq:pk_det_bridge}
\end{align}
up to terms depending on one site only, whose mixed derivative vanishes. Direct differentiation of the two representations yields
\begin{align}
 \operatorname{Tr}[G\,(\partial_iV)\,G\,(\partial_jV)]
 &=\operatorname{Tr}[\tau\,(\partial_it^{-1})
 \tau\,(\partial_jt^{-1})],\quad i\ne j,
 \label{eq:pk_vertex_bridge}
\end{align}
where the traces on the two sides are respectively in the electronic and site-scattering representations. This equality includes the local-state dressing and is the required \emph{direct Green-function--$T$-matrix connection}. Its spin-flip components, with the transverse rotational vertices and the same spin normalization on both sides, turn Eq.~(\ref{eq:pk_spectral}) into the original multiple-scattering result, Eq.~(19) of Ref.~\cite{ref18}:
\begin{align}
 \boxed{J_{ij}^{\rm LKG}=\frac{1}{4\pi}\operatorname{Im}
 \int_{-\infty}^{E_F}\!dE\,\operatorname{Tr}_L
 [\Delta_i^t\tau_{ij}^\uparrow\Delta_j^t\tau_{ji}^\downarrow].}
 \label{eq:pk_lkg_equality}
\end{align}
The inverse-$t$ vertex is $\Delta_i^t=(t_i^\uparrow)^{-1}-(t_i^\downarrow)^{-1}$, as defined above. In the common reference, the difference of the two inverse scattering-path operators is also exactly
\begin{align}
 \tau^\uparrow-\tau^\downarrow
 &=\tau^\uparrow[(\tau^\downarrow)^{-1}-(\tau^\uparrow)^{-1}]\tau^\downarrow
 =-\tau^\uparrow\Delta^t\tau^\downarrow.
 \label{eq:pk_tau_identity}
\end{align}
This last relation is the spin-channel identity used in Eqs.~(20)--(22) of Ref.~\cite{ref18}; it is a useful check, but the mixed-variation identity (\ref{eq:pk_vertex_bridge}) supplies the actual conversion between the two representations.

The PK and LKG expressions were obtained by different derivations: PK used perturbation theory for rotations of the local exchange fields, whereas LKG employed a multiple-scattering formulation. Both expressions describe the same second-order one-electron contribution evaluated at fixed potential. Although LKG also discussed the LFT, its first-order statement does not establish the corresponding identity for second derivatives of the self-consistent total energy. The equivalence of the PK and LKG exchange kernels follows instead from their common one-electron response and the transformation between orbital Green functions and scattering-path operators, with consistently normalized radial solutions and local vertices. This equivalence does not require the LFT and is distinct from the local-frame covariance of the rotational Hessian discussed in Sec.~\ref{sec:pt_repr}.\par
The coefficient $1/(4\pi)$ is the same as in Eq.~(\ref{eq:pk_spectral}), and both expressions refer to the historical ordered-pair convention in Eq.~(\ref{eq:pk_heis_original}). To use the half-sum convention of the main text, both PK and LKG coefficients must be doubled; this does not affect their equivalence.

The equivalence concerns the off-site frozen-potential rotational coefficient for the same local, collinear, nonrelativistic one-electron problem. It does not establish configuration-independent exchange interactions after self-consistent relaxation at arbitrary wave vector. The 1979 PK derivation also invokes its explicit short-range-orientation approximation in obtaining the Heisenberg form, a distinction from the algebraic equivalence of the resulting band and scattering kernels.

\section{Stationary Luttinger--Ward second variation}
\label{app:lw_variation}
For a fixed moment operator and source $h_B=-\widehat M B$, take the stationary functional
\begin{equation}
\mathcal F[G,B]=-\operatorname{Tr}\ln(-G^{-1})-\operatorname{Tr}[(G_{0,B}^{-1}-G^{-1})G]+\Phi[G],
\label{eq:lw_full_functional}
\end{equation}
with $G^{-1}=G_{0,B}^{-1}-\Sigma$, $\Sigma=\delta\Phi/\delta G$, and $\Gamma=\delta\Sigma/\delta G$. The trace includes Matsubara frequencies. For $G\to G+g$ and $h_B\to h_B+v$, the terms quadratic in $g$ and $v$, including the mixed source--response term, are
\begin{align}
\mathcal F_{\log}^{(2)}&=-\tfrac12\operatorname{Tr}(G^{-1}gG^{-1}g),\nonumber\\
\mathcal F_{\mathrm{sub}}^{(2)}&=\operatorname{Tr}(vg),\nonumber\\
\mathcal F_{\Phi}^{(2)}&=\tfrac12\operatorname{Tr}[g(\Gamma\cdot g)].\nonumber
\end{align}
The first-order terms in $g$ cancel by Dyson stationarity. Since $\delta G^{-1}=-G^{-1}gG^{-1}$, the terms above combine into Eq.~\eqref{eq:corr_lw_quadratic}. Variation with respect to $g$ gives $g=G(v+\Gamma\cdot g)G$ and reduces the stationary quadratic energy to $F^{(2)}=\tfrac12\operatorname{Tr}(vg)$. With $v=-\widehat M\delta B$ and $M=\operatorname{Tr}(\widehat M G)$, this yields $F_{BB}=-X$. The Legendre transform $E[M]=F[B]+BM$ then gives $E_{MM}=X^{-1}$. Conversion to angular coordinates additionally produces the geometric second-angular-derivative term in Eq.~\eqref{eq:orig103}. This proof assumes a differentiable, spin-consistent $\Phi$, a linear source, and an inverse on the allowed response subspace.

\section{Two-electron two-site test of stationary perturbation theory with nonlocal exchange}
\label{app:twosite}

\subsection{Functional and self-consistent antiferromagnetic state}
The two-site model provides an explicit comparison of the moment and field energy curvatures, including the local limit in which they differ. We test the stationary nonlocal theory using two sites $A,B$, one spatial orbital per site, and two electrons. In the spin basis $(A\uparrow,A\downarrow,B\uparrow,B\downarrow)$, we define
\begin{align}
 E[\gamma]&=\operatorname{Tr}(h_0\gamma)
 -\frac U4\bigl(|\mathbf M_A|^2+|\mathbf M_B|^2\bigr)\nonumber\\
 &\quad-\frac{K_{\mathrm{nl}}}{2}\operatorname{tr}_s(\gamma_{AB}\gamma_{BA}),\label{eq:appfunctional}\\
 h_0&=-t\sum_s\bigl(|As\rangle\langle Bs|\nonumber\\
 &\qquad\quad+|Bs\rangle\langle As|\bigr),\nonumber\\
 \mathbf M_i&=\operatorname{tr}_s(\bm{\sigma}\gamma_{ii}).\nonumber
\end{align}
The density matrix $\gamma$ describes two occupied one-electron states and satisfies $\gamma^2=\gamma$. We use this rotationally invariant model to illustrate nonlocal exchange, without identifying it with a particular exchange--correlation approximation. The resulting one-electron Hamiltonian contains an exchange term connecting sites $A$ and $B$:
\begin{equation}
 \begin{aligned}
 h_{ii}&=-\frac U2\mathbf M_i\cdot\bm{\sigma},\\
 h_{AB}&=-t\mathbf1-\frac{K_{\mathrm{nl}}}{2}\gamma_{AB},\qquad h_{BA}=h_{AB}^{\dagger}.
 \end{aligned}
 \label{eq:appH}
\end{equation}
For $t>0$ and $d\equiv 2U-K_{\mathrm{nl}}>4t$, we consider the collinear AF determinant with occupied orbitals
\begin{align}
 |h_1\rangle&=c|A\uparrow\rangle+a|B\uparrow\rangle,\nonumber\\
 |h_2\rangle&=a|A\downarrow\rangle+c|B\downarrow\rangle,\qquad a^2+c^2=1.
 \label{eq:appoccupied}
\end{align}
Its local moments are $\mathbf M_A=m\hat{\mathbf z}$ and $\mathbf M_B=-m\hat{\mathbf z}$, where $m=c^2-a^2>0$. In this AF state, the intersite density matrix is $\gamma_{AB}=ac\mathbf1$. The nonlocal exchange term therefore changes the effective hopping by $-K_{\mathrm{nl}}ac\mathbf1/2$. It is spin independent in the collinear reference state, but its induced response to nonuniform spin rotations is spin dependent. The total energy is
\begin{equation}
 E_{\rm AF}(a)=-4tac-\frac U2(c^2-a^2)^2-K_{\mathrm{nl}}a^2c^2.
 \label{eq:appEaf}
\end{equation}
Setting $x=2ac$ reduces this expression to
\begin{equation}
 \begin{aligned}
 E_{\rm AF}(x)&=-\frac U2-2tx+\frac d4x^2,\\
 x_*&=\frac{4t}{d},\qquad m=\sqrt{1-\frac{16t^2}{d^2}}.
 \end{aligned}
 \label{eq:appselfcons}
\end{equation}
At the self-consistent minimum, the AF energy is $E_{\rm AF}=-U/2-4t^2/d$. The fully polarized FM determinant has $E_{\rm FM}=-U/2$, so their difference is $-4t^2/d$. This finite-angle energy difference is \emph{not} used to define the local pair exchange. In this particular model the two definitions nevertheless coincide: a bilinear pair energy $-J_{AB}\mathbf e_A\cdot\mathbf e_B$ gives $E_{\rm FM}-E_{\rm AF}=-2J_{AB}$, and the difference $4t^2/d$ yields the same $J_{AB}=-2t^2/d$ obtained below from the local change of magnetic energy. The model therefore illustrates the stationary nonlocal formalism but does not discriminate between the two definitions.

\subsection{Transverse particle--hole Hessian and constrained exchange}
The unoccupied eigenstates can be chosen as
\begin{equation}
 |p_1\rangle=-a|A\uparrow\rangle+c|B\uparrow\rangle,\qquad
 |p_2\rangle=c|A\downarrow\rangle-a|B\downarrow\rangle.
 \label{eq:appempty}
\end{equation}
A small transverse rotation mixes occupied and empty states. For one real transverse direction, we write the first-order change in the density matrix as
\begin{align}
 \delta\gamma={}&z_1(|p_2\rangle\langle h_1|+\mathrm{h.c.})\nonumber\\
 &+z_2(|p_1\rangle\langle h_2|+\mathrm{h.c.}).
 \label{eq:apptrans}
\end{align}
The projected transverse moments are
\begin{equation}
 \delta M_A^x=2(c^2z_1-a^2z_2),\qquad
 \delta M_B^x=2(-a^2z_1+c^2z_2).
 \label{eq:appmoments}
\end{equation}
To make the quadratic expansion explicit, write
$\gamma=\gamma_0+\gamma_1+\gamma_2+O(z^3)$, where
$\gamma_1=[\Omega,\gamma_0]=\delta\gamma$ and
$\gamma_2=\tfrac12[\Omega,[\Omega,\gamma_0]]$.
Here $\Omega$ mixes only the occupied and empty orbitals in
Eq.~\eqref{eq:apptrans}, and $\gamma_2$ preserves the idempotency
of the density matrix to second order. Let
$\mathbf M_i^{(n)}=\operatorname{tr}_s(\bm\sigma\gamma_{n,ii})$.
Substitution into Eq.~\eqref{eq:appfunctional} gives the quadratic
energy before simplification:
\begin{align}
 E_\perp^{(2)}={}&\operatorname{Tr}(h_0\gamma_2)
 -\frac U2\sum_i\mathbf M_i^{(0)}\cdot\mathbf M_i^{(2)}\nonumber\\
 &-\frac U4\sum_i|\mathbf M_i^{(1)}|^2\nonumber\\
 &-\frac{K_{\mathrm{nl}}}{2}\operatorname{tr}_s\bigl(
   \gamma_{1,AB}\gamma_{1,BA}\nonumber\\
 &\qquad+\gamma_{0,AB}\gamma_{2,BA}
  +\gamma_{2,AB}\gamma_{0,BA}\bigr).
 \label{eq:appPTexpanded}
\end{align}
The first term contains the second-order change in hopping energy.
The next two terms account for the change of the local moments,
including their transverse components. The last line retains both
the first-order change of the intersite density matrix and its
second-order change; neither can be discarded for a nonlocal
magnetic interaction. Inserting the orbitals in
Eqs.~\eqref{eq:appoccupied} and \eqref{eq:appempty} and using the
stationary relation $2ac=4t/d$ reduces Eq.~\eqref{eq:appPTexpanded} to
\begin{equation}
 E_{\perp}^{(2)}=d\,a^2c^2(z_1+z_2)^2
 =\frac12\begin{pmatrix}z_1&z_2\end{pmatrix}
 \underbrace{2da^2c^2\begin{pmatrix}1&1\\1&1\end{pmatrix}}_{A_\perp}
 \begin{pmatrix}z_1\\z_2\end{pmatrix}.
 \label{eq:appPT}
\end{equation}
The generator $\Omega$ is anti-Hermitian, so this expansion is equivalent to $\gamma=e^{\Omega}\gamma_0e^{-\Omega}$ and keeps the occupied orbitals orthonormal. All interaction terms are included; no frozen-band-energy approximation is used. A uniform spin rotation costs no energy: the Hessian has a zero mode $(z_1,z_2)\propto(1,-1)$ and a second eigenvalue $4da^2c^2$. The other transverse polarization has the corresponding zero mode.

For the AF unit-vector angular coordinates, $\delta M_A^x=m\theta_A$ and $\delta M_B^x=-m\theta_B$. Equation~(\ref{eq:appmoments}) implies
\begin{equation}
 z_1+z_2=\frac{\theta_A-\theta_B}{2},\qquad
 E_{\rm constr}^{(2)}=\frac{t^2}{d}(\theta_A-\theta_B)^2.
 \label{eq:appconstr}
\end{equation}
The expansion of $-J_{AB}\mathbf e_A\cdot\mathbf e_B$ around antiparallel moments is $-J_{AB}(\theta_A-\theta_B)^2/2$, giving
\begin{equation}
 \boxed{J_{AB}^{M}=-\frac{2t^2}{2U-K_{\mathrm{nl}}}},\qquad
 J_{AB}^{M}(K_{\mathrm{nl}}=0)=-\frac{t^2}{U}.
 \label{eq:appJ}
\end{equation}
This coefficient describes small rotations about the stationary AF state; it need not describe other magnetic configurations.

\subsection{Prescribed-field exchange in the local two-site model}
\label{app:twosite_frozen}
For $K_{\mathrm{nl}}=0$, the same model permits an independent calculation of the field curvature. At the self-consistent AF reference, $m^2=1-4t^2/U^2$ and the local exchange-field magnitude is $b=Um/2$. Keeping $b$ fixed, we rotate the two local exchange fields through a relative departure $\delta$ from antiparallel alignment. The frozen one-electron Hamiltonian is
\begin{equation}
 h^B(\delta)=-t\tau_x-b\bigl(P_A\mathbf e_A+P_B\mathbf e_B\bigr)\cdot\bm\sigma,
 \qquad \mathbf e_A\cdot\mathbf e_B=-\cos\delta,
 \label{eq:appfrozenH}
\end{equation}
where $P_A$ and $P_B$ project onto the two sites and $\tau_x$ acts in site space. Equation~\eqref{eq:appfrozenH} rotates prescribed local exchange fields at fixed magnitude, rather than constraining the directions of self-consistent magnetic moments. The electronic potential is not readjusted to enforce specified moment directions. Its energy curvature therefore describes exchange-field rotations, not the fully relaxed energy curvature for moment rotations. This distinction remains even for a purely local interaction. Its four eigenvalues are $\pm\sqrt{t^2+b^2\pm 2tb\sin(\delta/2)}$. The two negative-energy levels are occupied. With $C=t^2+b^2$, their sum gives
\begin{align}
 E_{\rm band}^B(\delta)-E_{\rm band}^B(0)
 &=\frac{t^2b^2}{4C^{3/2}}\delta^2+O(\delta^4)\nonumber\\
 &=\frac{t^2m^2}{2U}\delta^2+O(\delta^4),
 \label{eq:appfrozenE}
\end{align}
where the second line uses $C=U^2/4$ at self-consistency. The frozen local interaction double-counting term is unchanged by these rotations. Matching the quadratic energy to $-J_{AB}^B\delta^2/2$, with the same pair-counting and angular convention as Eq.~\eqref{eq:appJ}, yields
\begin{equation}
 \boxed{J_{AB}^B=m^2J_{AB}^M=-\frac{t^2}{U}\left(1-\frac{4t^2}{U^2}\right)}\qquad(K_{\mathrm{nl}}=0).
 \label{eq:appfrozenJ}
\end{equation}
Thus the field and moment pair coefficients differ even for a local, spin-rotation-invariant interaction. Their ratio $J_{AB}^B/J_{AB}^M=1-4t^2/U^2$ approaches unity as $t/U\to0$, the strong-moment limit. This finite-bond AF result is distinct from the equality of the leading $q^2$ curvatures in the scalar ferromagnetic model of Sec.~\ref{sec:pt_fm}.

\subsection{Independent finite-angle energy and stability conditions}
We check the result independently by rotating the occupied orbitals in Eq.~(\ref{eq:appoccupied}) so that the two AF moments turn toward one another by a relative angle $\delta$. The local moment magnitudes and the bond invariant $\operatorname{tr}_s(\gamma_{AB}\gamma_{BA})$ remain fixed under this path. The hopping energy changes according to
\begin{align}
 E_{\rm path}(\delta)&=-4tac\cos(\delta/2)
 -\frac U2m^2-K_{\mathrm{nl}}a^2c^2,\nonumber\\
 E_{\rm path}(\delta)-E_{\rm AF}
 &=4tac[1-\cos(\delta/2)]
 =\frac{t^2}{d}\delta^2+O(\delta^4).
 \label{eq:apppath}
\end{align}
This rotated state is a variational path, not necessarily the lowest constrained-energy state at finite $\delta$. For small rotations, however, the two transverse amplitudes in Eq.~(\ref{eq:apptrans}) are fixed by the moment directions. The other particle--hole channels describe longitudinal or charge changes to first order and do not alter the quadratic transverse energy when the AF reference is stable. The finite-angle result thus checks the constrained local change of magnetic energy in Eq.~(\ref{eq:appconstr}).

At $(t,U,K_{\mathrm{nl}})=(1,4,1)$, numerical differentiation of the full eight-dimensional real particle--hole Hessian gives two Goldstone zero modes and six positive eigenvalues. This confirms local AF stability for these parameters, but not for the entire range $d>4t$ in which the AF state exists.

The model gives a $K_{\mathrm{nl}}$-dependent constrained exchange second derivative. Stationary perturbation theory agrees with the finite-angle expansion and passes the Goldstone and stability tests. Its two transverse amplitudes are fixed by the moment constraints, so it does not independently validate relaxation among extra unconstrained transverse channels; that would require additional orbitals or sites. In particular, Eq.~(\ref{eq:appJ}) must not be extrapolated to a general nonlocal functional or interpreted as the prescribed-field pair coefficient of Sec.~\ref{sec:nonlocalstationary} [Eq.~\eqref{eq:nlGF}].

\begin{acknowledgments}
Work at Ames National Laboratory was supported by the U.S. Department of Energy (DOE), Office of Science, Basic Energy Sciences, Materials Sciences and Engineering Division. Ames National Laboratory is operated for the U.S. DOE by Iowa State University under Contract No. DE-AC02-07CH11358.
\end{acknowledgments}

\section*{Data Availability}
The analytical results and the numerical values reported in this article are contained in the manuscript. No external datasets were used.

\end{document}